\documentclass[fleqn,usenatbib]{mnras}
\usepackage{fix-cm}
\usepackage{newtxtext,newtxmath}
\usepackage[T1]{fontenc}
\usepackage{lineno}

\DeclareRobustCommand{\VAN}[3]{#2}
\let\VANthebibliography\thebibliography
\def\thebibliography{\DeclareRobustCommand{\VAN}[3]{##3}\VANthebibliography}

\usepackage{graphicx}	% Including figure files
\usepackage{amsmath}	% Advanced maths commands

\usepackage{float}
\usepackage{color,xcolor}
\usepackage{caption}
\usepackage{subcaption}

\title[Multi-messenger modeling of the Monogem pulsar halo]{Multi-messenger modeling of the Monogem pulsar halo}

\author[Li et al.]{
Youyou Li,$^{1}$\thanks{E-mail: y.li4@uva.nl}
Oscar Macias,$^{2,1}$\thanks{macias@sfsu.edu}
Shin'ichiro Ando,$^{1,3}$ and
Jacco Vink$^{1}$
\\
$^{1}$GRAPPA Institute, University of Amsterdam, 1098 XH Amsterdam, The Netherland\\
$^{2}$Department of Physics and Astronomy, San Francisco State University, San Francisco, California 94132, USA\\
$^{3}$Kavli Institute for the Physics and Mathematics of the Universe, University of Tokyo, Chiba 277-8583, Japan
}

\date{Accepted 2025 February 24. Received 2025 February 24; in original form 2024 July 9}

\pubyear{2025}

\begin{document}
\label{firstpage}
\pagerange{\pageref{firstpage}--\pageref{lastpage}}
\maketitle

\begin{abstract}
The High-Altitude Water Cherenkov Telescope (HAWC) has detected TeV halos associated with two nearby pulsars/pulsar wind nebulae (PWN) --- Geminga and B0656+14. These TeV halos extend up to tens of pc from the central accelerators, indicating that the diffusion of
ultrarelativistic electrons and positrons in the interstellar medium has been suppressed by two orders of magnitude. Although Geminga and B0656+14 are at similar distances and in the same field of view, they have distinct histories. Notably, B0656+14 probably still resides within its parent supernova remnant, the Monogem Ring, which can be observed in X-rays. In this work, we perform high-resolution simulations of the propagation and emission of relativistic lepton pairs around B0656+14 using a two-zone diffusion model using the GALPROP numerical code. We compared the predicted inverse-Compton spectrum to the observations made by HAWC and Fermi-LAT and found physically plausible model parameters that resulted in a good fit to the data. Additionally, we estimated the contribution of this TeV-halo to the positron flux observed on Earth and found it to be smaller than 10\% of the measured flux. We conclude that future observations of the TeV halo and its synchrotron emission counterpart in radio and X-ray frequencies will be crucial to distinguish between various possible models.
\end{abstract}


%%%%%%%%%%%%%%%%%%%%%%%%%%%%%%%%%%%%%%%%%%%%%%%%%%

%%%%%%%%%%%%%%%%% BODY OF PAPER %%%%%%%%%%%%%%%%%%

\section{Introduction}\label{section: intro}
B0656+14 is a middle-aged pulsar with a characteristic age of approximately 110,000 years. It is located about 288~pc away from Earth, as determined through parallax measurements~\citep{Brisken_2003}.
The estimated pulsar age and distance is consistent with those of the ``Monogem Ring'' supernova remnant, a ring-shaped structure visible in X-rays with an apparent extension of 25$^{\circ}$, implying that they likely share a common origin \citep{Thorsett_2003}. We refer to B0656+14 as the ``Monogem pulsar'', hereafter. Monogem, first discovered as a radio pulsar \citep{Manchester_1978}, has been extensively studied at all wavelengths from radio to gamma rays (e.g., \citealt{Zharikov_2021, Durant_2011, Schwope_2022, Abdo_2010}).

\begin{figure}
     \centering
     \includegraphics[width=0.45\textwidth]{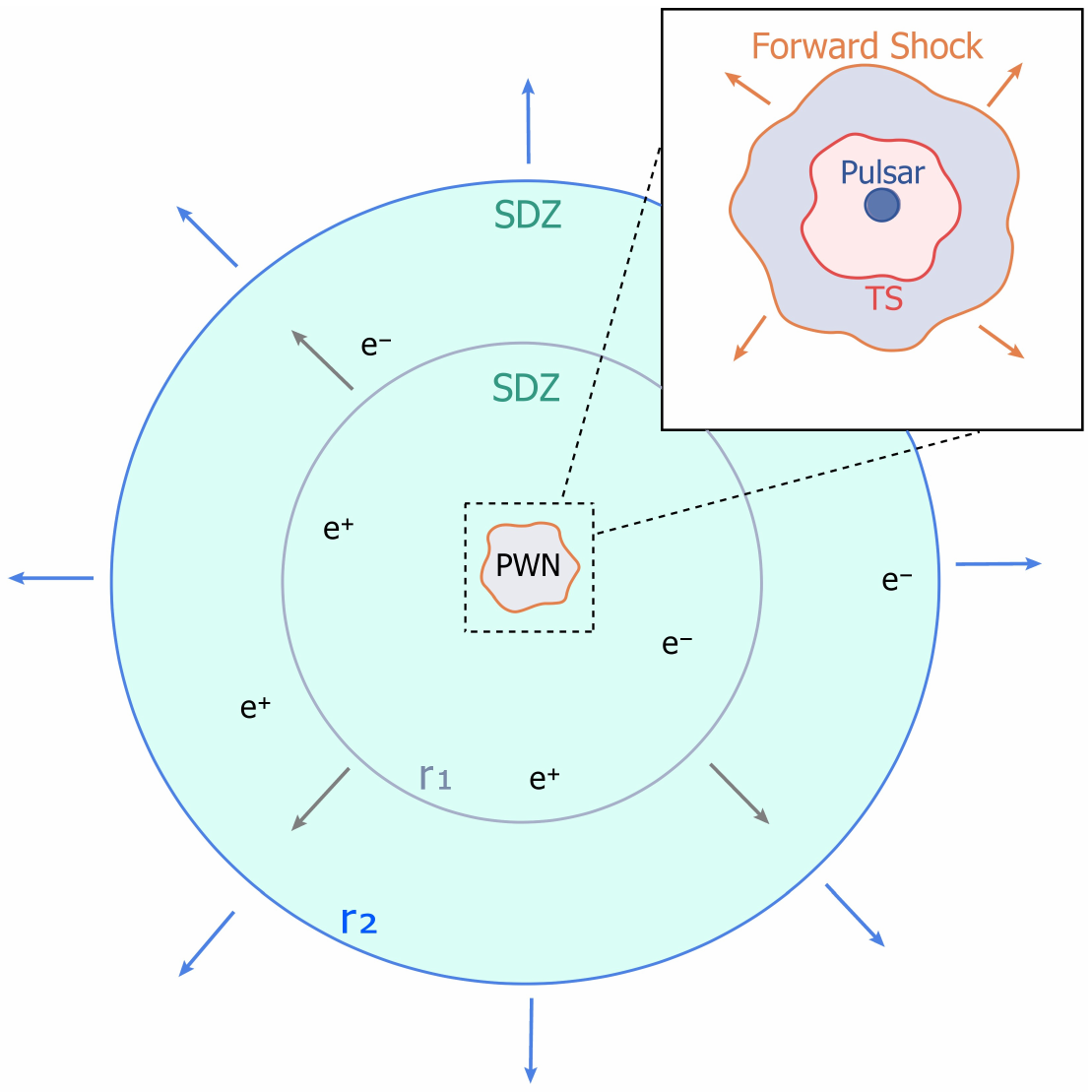}
     \caption{Illustration of PWN two-zone diffusion model. TS (termination shock) marks the acceleration site of the electron positrons escaped from the PWN. $r_{1}$ is the transition radius, and $r_{2}$ is the radius of the SDZ.}  
     \label{fig: SDZ_illustration}
\end{figure}

\hfill\break
Monogem has a rotation period of $P\approx 385~\mathrm{ms}$,
and a spin-down power $(\dot{E})$ of $3.8 \times 10^{34}$ erg $\mathrm{s}^{-1}$. It has a surface magnetic field $(B_{\text {surf}})$ of $4.65 \times 10^{12}\; \mathrm{G}$. Pulsars like Monogem emit highly energetic winds of electrons and positrons escaping along open magnetic field lines from the magnetosphere. The wind ends in a termination shock at the radius where the ram pressure of the wind equals the pressure of the medium surrounding the pulsar (cf. Fig.~\ref{fig: SDZ_illustration}). These shocks create pulsar wind nebulae (PWNe), which can be observed through the non-thermal radiation emitted by the injected particles. The brightness of PWNe depends on the pulsar's spin-down power which declines with pulsar age. So older pulsars are typically more challenging to detect. Interestingly, the PWN associated with Monogem is one of the few PWNe that have been identified. It was detected in between $\sim 0.005\text{--}0.2$ pc radius around Monogem using Chandra X-ray data, showing a roughly round morphology \citep{Birzan_2016}. The X-rays are thought to be caused by synchrotron radiation of $e^{\pm}$ in the downstream region of the termination shock \citep{Birzan_2016}, which are expected to produce very-high-energy (VHE) counterparts at TeV energies through inverse-Compton (IC) scattering of the ambient photon field \citep{kargaltsev2013gamma}.

\hfill\break
The first evidence of TeV emission around the Monogem pulsar was discovered through Milagro observations \citep{Abdo_2009}. Subsequently, more details about the TeV emission were revealed by HAWC with the detection of TeV halos around the PWNe of both the Monogem pulsar and another middle-aged pulsar known as Geminga \citep{HAWC_2017}. Pulsar halos \citep{Linden_2017} are emission envelopes stretching up to tens of parsecs, which are purported to have been produced by a population of relativistic $e^{\pm}$ that have diffused out of the PWNe into the ISM. So far, these haloes have only been identified in gamma-ray emission above a few TeV. Recent studies \citep{HAWC_2017, Profumo:2018fmz,hooper_2017,  Di_Mauro_2019,Manconi_2020} suggest that to explain the vast extension of these TeV halos, it is necessary to have a reduction in the diffusion coefficient by approximately two orders of magnitude over distances ranging from tens to hundreds of parsecs around the parent pulsar wind nebulae. An alternative model suggests that the observations can be explained by considering the transition between the quasi-ballistic regime and the diffusion regime in the propagation of electrons and positrons after they leave the acceleration site \citep{Recchia_2021}.
However, this model has been critiqued for requiring an acceleration efficiency exceeding 100\% of the pulsar's spin-down power \citep{Bao_2022}. Therefore, the slower diffusion process of cosmic rays (CRs) in the nearby ISM of the parent PWN remains the most plausible explanation for the TeV halos.

\hfill\break
The biggest challenge in understanding TeV halos is to explain why the diffusion of particles with energies in the multi-TeV range is suppressed around the source for hundreds to thousands of years \citep{Recchia_2021, Martin_2022, López-Coto_2018}. Some studies have put forward a well-motivated theory that suggests the origin of magnetic turbulence is environmental, caused by the host supernova remnant or a pre-existing feature in the ISM in which the TeV-halo develops \citep{Fang_2019}. On the other hand, independent studies have proposed that the self-induced resonant and non-resonant streaming instability caused by the propagation of the electrons and positrons may lead to strong self-confinement in the vicinity of the sources reaching 50--100 pc from the PWN \citep{Malkov_2013, Nava_2016, Evoli_2018, Nava_2019, Schroer_2021}.

\hfill\break
These hypotheses may leave different imprints on the halo emission morphology. In the case of electron/positron self-confinement, the diffusion suppression region is expected to evolve dynamically with the pulsar. In the presence of environmental turbulence, the confinement region is expected to share the center of the parent SNR and evolve in conjugation with the expansion of the SNR's shock wave and its interaction with the pulsar wind. These effects have been demonstrated phenomenologically by \citet{Jóhannesson_2019} and \citet{Di_Mauro_2019} for the case of the Geminga pulsar. The slow diffusion phenomena around Geminga pulsar and Monogem have been explored analytically by \citet{Martin_2022} and \citet{Schroer_2023}, but both have assumed a static system, leaving the impact of the proper motion of the pulsar and origin of the slow diffusion out of scope. 

\hfill\break
In this paper, we investigate the propagation of CRs in the Monogem PWN's surrounding interstellar medium (ISM). We employ a numerical approach on a two-zone diffusion model to achieve this. The propagation of CRs and non-thermal emissions are calculated using the GALPROP numerical code \citep[v57; ][]{Porter:2021tlr}. This package enables accurate forecasts of the gamma-ray emission morphology and
radiation spectra by using advanced modeling of the interstellar radiation field \citep{Porter_2017}, spatial dependent magnetic field, and three-dimensional interstellar gas \citep{Johannesson:2018bit}. It also allows a non-equidistant spatial grid, which provides exceptionally high spatial resolution around the astrophysical accelerator. We compare our simulated non-thermal maps with HAWC observations and Fermi-LAT flux upper limits in the 8--40 TeV and 10--1000 GeV energy range, respectively. We explore how the pulsar's age, injection spectra, slow diffusion zone size, magnetic field, and proper motion of the Monogem pulsar impact the IC and synchrotron emission properties. In the near future, advanced gamma-ray detectors such as the Cherenkov Telescope Array Observatory (CTAO) could verify our predictions of the spatial shape and spectrum of Monogem, as analyzing Monogem using the conventional ON/OFF method may prove difficult due to the large extension of its emission and the energy-dependent nature of its gamma-ray spatial morphology \citep{HESS_2023,Eckner_2023}. In addition, we offer hints about possible future searches of the synchrotron halo emission across radio to X-ray frequencies. Finally, utilizing models that align with the electromagnetic observations across multiple wavelengths, we put forth projections for the Monogem pulsar's contribution to the positron flux that AMS-2 has measured.

\hfill\break
This paper is structured as follows: In section \ref{sec:modeling}, we introduce the  properties of Monogem pulsar and the modeling of the injection and transportation of leptonic particles from the pulsar. In section \ref{sec:results}, we compare the diffusive synchrotron and IC emission around Monogem to the HAWC and Fermi-LAT data. We also compare the expected positron flux from Monogem to the positron flux measured by AMS-2. The injection and propagation properties of CRs around Monogem pulsar are discussed.
\section{Modeling of Non-thermal Emission from the Monogem Pulsar Halo}\label{sec:modeling}
\begin{table}
    \centering
    \resizebox{0.48\textwidth}{!}{ % Adjust to fit in one column width
    \begin{tabular}{|p{3.7cm}|p{3cm}|p{1.3cm}|}
    \hline \hline 
    Parameter & Values & Default Value \\ \hline \hline
    Source & 2HWC J0700+143 & - \\
    Associated pulsar & B0656+14 (Monogem pulsar) & - \\
    Present period $(P)$ & $384.94$ ms & - \\
    Period derivative $(\dot{P})$ & $5.5\times10^{-14} \mathrm{~s} \mathrm{~s}^{-1}$ & - \\
    Initial spin-down power $(\dot{E}_0)$ & $1.84 \times 10^{35}\; \mathrm{erg}\; \mathrm{s}^{-1}$ & - \\
    Characteristic age $(\tau_{\rm c})$ & $110\; \mathrm{kyr}$ & - \\
    Current age $(t_{\rm age})$ & $11\; \mathrm{kyr}$, $55\; \mathrm{kyr}$, $99\; \mathrm{kyr}$ & $99\; \mathrm{kyr}$ \\
    Distance $(d)$ & $288\; \mathrm{pc}$ & - \\
    Galactic coordinates $(l, b)$ & $(201.1^{\circ}, 8.3^{\circ})$ & - \\
    Galactocentric coordinates $(X, Y, Z)$ & $(-8.77, -0.10, 0.04)\; \mathrm{kpc}$ & - \\
    \hline
    Diffusion coefficient at $100\; \mathrm{TV}$ in ISM $(D_{\mathrm{ISM}})$ & $1.56\times 10^{30}$ cm$^2$ s$^{-1}$ & - \\
    Diffusion bubble core $(r_1)$ & $25\; \mathrm{pc}$, $50\; \mathrm{pc}$ & $50\; \mathrm{pc}$ \\
    Diffusion bubble radius $(r_2)$ & $45\; \mathrm{pc}$, $70\; \mathrm{pc}$ & $70\; \mathrm{pc}$ \\
    Energy break $(E_{\mathrm{b}})$ &  $100\; \mathrm{GeV}$ & $100\; \mathrm{GeV}$ \\
    Smoothness $(s)$ & $0.5$ & - \\
    Low-energy spectral index $(\gamma_0)$ &  $1.5$ & $1.5$ \\
    High-energy spectral index $(\gamma_1)$ & $1.8$--$2.5$ & $2.2$ \\
    Magnetic field strength $(B)$ & $3\mu$G, $5\mu$G, $10\mu$G & $3\mu$G \\
    %Spin-down power efficiency $(\eta)$ & $3$--$16\%$ & - \\
    \hline
    \end{tabular}
    }
    \caption{Parameters used to model the pulsar and the slow diffusion zone. When the the parameters are not stated otherwise, default values are used.}
    \label{tab:Pulsar_Model_parameters}
\end{table}

\subsection{Particle Injection from Monogem}
\label{sec:injection} 
To better understand the extended non-thermal emissions surrounding the Monogem pulsar, we must first determine the injection properties of the pairs responsible for these emissions. According to recent HAWC results, the TeV halo around Monogem is caused by the IC emission of high-energy $e^{\pm}$ particles that have escaped from the PWN and made their way into the ISM. These $e^{\pm}$ particles are accelerated at the PWN, which is powered by the pulsar's spin-down. Currently, the spin-down power of the pulsar is estimated to be $\dot{E}= 3.8\times10^{34}$ erg s$^{-1}$, as reported by \cite{Manchester_2005}. Pulsar wind material, which consists mainly of $e^{\pm}$, enters the termination shock of the pulsar wind and the particles will obtain a nonthermal energy distribution as the result of diffusive shock acceleration at the shock, or due to acceleration by magnetic field reconnection processes in the wind itself. Multi-wavelength studies of individual PWNe suggest that the $e^{\pm}$ injection spectrum of PWNe follows a broken power-law distribution, with a low-energy index in the range $[1.0,2.0]$, a high-energy index in the range $[2.0, 2.8]$, and a spectral break occurring at $[10, 100]$ GeV \citep{Bucciantini_2010,Martin_2022,MARTIN_J_2022,TORRES2014}.

\hfill\break
In our simulation, we do not consider the spatial extent of the PWN, because it only covers a radius of approximately 0.2 pc. This is much smaller than the extent of the halo emission we are interested in. Additionally, we assume that electrons and positrons undergo the same acceleration and propagation process. The pairs are injected into the ISM from the pulsar's location in an isotropic manner, following a smooth, broken power-law spectrum, described by the expression \citep{Jóhannesson_2019}:
\begin{equation}
    \frac{dn_{e^{\pm}}}{dE}\propto E^{-\gamma_{0}}\left[1+\left(\frac{E}{E_{\rm{b}}}\right)^{\frac{\gamma_{1}-\gamma_{0}}{s}}\right]^{-s},
\end{equation}
where $n_{e^{\pm}}$ is the number density of the $e^{\pm}$, and $E$ is the particle kinetic energy. The power-law indices $\gamma_{0}$ and $\gamma_{1}$ are the spectral indices at low and high energies, respectively. As the low energy index $\gamma_{0}$ is poorly constrained, we adopt a mean value of $\gamma_{0}$=1.5, as indicated by observations of other PWNe. We explored scenarios where $\gamma_{1}=$ 1.8, 2.0, 2.2, 2.5. Notably, while $\gamma_{1}=$ 1.8 falls out of the range of typically suggested high energy indices of PWNe, the HAWC spectral shape of the Monogem pulsar halo indicates such a hard injection spectrum is possible, thus we include it in our calculation. We inject electrons from 100 MeV to an energy cutoff of 1 PeV, with a break energy at $E_{\rm{b}}=100$ GeV by default.

\hfill\break
The normalization of the injected spectrum is a free parameter to ensure that the total luminosity of the injected pairs $L_{e^{\pm}}$ at time $t$ remains a fraction of the spin-down power $\dot{E} (t)$ of the pulsar \citep{Pacini_1973}: 
\begin{equation}    
   L_{e^{\pm}}(t)=\eta\dot{E_{0}}\left(1+\frac{t}{\tau_{0}}\right)^{-\frac{n+1}{n-1}},
\end{equation}
where $\dot{E_{0}}$ is the initial spin-down power of the pulsar, and $\eta$ is the fraction of the pulsar spin-down power that is converted into the luminosity of injected pairs. The $n$ in the index is the braking index of a pulsar. Until now, only a few pulsars have their braking index accurately measured. Previous studies show $n=15$ for Monogem pulsar, but the error bar spans $\sim 120$ in value, leaving the actual value ambiguous \citep{Johnston_1999}.  We adopt the pulsar braking index corresponding to the magnetic-dipole model, thus $n=3$, which is also shown to yield a good agreement between the pulsar characteristic age and the kinetic age for middle-aged pulsars \citep{Bailes_1989, Lorimer_1997}. The initial spin-down time-scale of the pulsar $\tau_{0}$ is connected with the current characteristic age $\tau_{\rm c}$ of the pulsar by the following expression: 
\begin{equation}\label{tau_0}
    \tau_{0}\equiv \frac{P_{0}}{(n-1)\dot{P_{0}}}=\frac{2\tau_{\rm c}}{(n-1)}-t_{\rm age}.
\end{equation}
For a braking index of $n=3$, equation~(\ref{tau_0}) becomes $\tau_{0}=\tau_{\rm c}-t_{\rm age}$. It is worth noting that the true age of a pulsar often has a considerable discrepancy with its characteristic age \citep{Brisken_2003_pulsarage}.

\hfill\break
One of the key arguments that the Monogem Ring and the Monogem pulsar have the same origin is that the pulsar characteristic age agrees with the estimated SNR age based on Sedov modeling \citep{Thorsett_2003}. This only holds if: 1) the characteristic age of Monogem pulsar is a representation of the true age of the pulsar, 2) the Monogem Ring's expansion has been governed by Sedov expansion up to the present, which, we note, is unlikely since this phase typically lasts only up to $\sim$20~kyr  \footnote{During the Sedov phase the explosion energy is conserved inside the shell, and the radius evolves as $R\propto t^{2/5}$. After $\sim 2\times 10^4$~yr, typically when the velocity $\lesssim 200~{\rm km/s}$, the remnant enters the radiative phase and $R\propto t^{1/4}$.}, compared to $\sim 110$~kyr of the characteristic pulsar age. Searching for evidence of the pulsar's true age in non-thermal diffused emissions can provide insight into the origin of the Monogem pulsar. In this study, we investigate how the initial timescale affects the diffuse emission by testing three different scenarios where $\tau_{0}$ is set to 0.1$\tau_{\rm c}$, 0.5$\tau_{\rm c}$ and 0.9$\tau_{\rm c}$. In the first scenario, it is assumed that the actual age of the pulsar is close to its characteristic timescale (which suggests that the pulsar had much higher spin-down power in the past).
\subsection{Slow Diffusion Zone}\label{SDZ} 
We assume the diffusion coefficient within a confined region around Monogem is lower than in the general ISM. This assumption is based on recent studies \citep[e.g.,][]{Profumo:2018fmz, J_hannesson_2019} of similar systems considering the Galactic magnetic field. This confined region, referred to as the ``slower diffusion zone'' (SDZ), is illustrated in Figure \ref{fig: SDZ_illustration}. The spatial dependence of the diffusion coefficient is then determined by
\begin{equation}
    D=\left(\frac{E}{E_{0}}\right)^{\delta}\times 
        \begin{cases}
            D_{\rm SDZ}, & r<r_{1},\\
            D_{\rm SDZ} \left( \frac{D_{\rm ISM}}{D_{\rm SDZ}}\right)^{(r-r_{1})/(r_{2}-r_{1})}, & r_{1}\leq r \leq r_{2},\\
            D_{\rm{ISM}}, & r>r_{2}.
        \end{cases}
\end{equation}
where the diffusion coefficient has a power law energy dependence with slope $\delta$. We adopt $\delta=0.35$, which leads to a Kolmogorov diffusion.\\
\hfill\break
The transport of particles within a bubble (of radius of $r_{1}$) of slow diffusion surrounding the pulsar is determined by the coefficient $D_{\rm SDZ}$. The diffusion coefficient gradually changes from $D_{\rm SDZ}$ to $D_{\rm ISM}$ between the transitional radius $r_{1}$ and the outer radius $r_{2}$, where $D_{\rm ISM}$ is the average diffusion coefficient in the general interstellar medium. The value of $D_{\rm SDZ}$ is determined by equating the diffusion length of $\sim$100 TeV particles to the extension of the $\sim$ 20~TeV gamma-ray halo observed by HAWC, which is 24.18~pc \citep{HAWC_2017}: 
\begin{equation}\label{eq: diffusion length}
    R_{\rm{diff}}=\sqrt{4D_{100\rm{TV}}min\{\tau_{\rm{cool}},\tau_{\rm{inj}}\}}.
\end{equation}
Here, $\tau_{\rm{inj}}$ is the injection time scale, determined by the age of the PWN. The cooling time $\tau_{\rm{cool}}$ is the energy loss time due to synchrotron and IC emission, therefore, it depends on the assumption of the magnetic field. It is expressed as: 
\begin{equation}\label{eq:cooling_time}
    \tau_{\rm cool}=\frac{4m_{\rm{e}}c^{2}}{3c\sigma_{T}\gamma}\left(U_{B}+
            \sum_i \frac{U_{\text{ph}, i}}{(1+4\gamma\varepsilon_{0, i})^{3/2}}\right)^{-1},
\end{equation}
where, $\sigma_{T}$ is the Thompson scattering cross section, $U_{B}$ and $U_{\rm{ph}}$ are the energy density of the magnetic field and the photon field respectively, and $\varepsilon_{0}$ is the ratio of average photon energy to the electron rest mass energy. When electrons/positrons have a Lorentz factor $\gamma$ greater than $\gamma_{\rm{KN}} \equiv \frac{1}{4\varepsilon_{0}}$, the Klein-Nishina (KN) effect becomes significant, suppressing their cooling rate. We consider the following components of interstellar photon fields: optical ($U_{\rm opt}=0.56$~eV~cm$^{-3}$, T$_{\rm opt}=5000$~K), infrared ($U_{\rm IR}=0.41$~eV~cm$^{-3}$, T$_{\rm IR}=20$~K), and the Cosmic Microwave Background ($U_{\rm CMB}=0.26$~eV~cm$^{-3}$, T$_{\rm CMB}=2.7$~K) \citep{john2023}. The cooling time for 100~TeV electron/positrons assuming a 3~$\mu$G magnetic field, is approximately 10~kyr. Thus, according to Eq. \ref{eq: diffusion length}, the diffusion length of the 100~TeV electrons/positrons is determined by the cooling time scale instead of the injection time scale for all pulsar ages tested (listed in Table \ref{tab:Pulsar_Model_parameters}).
\hfill\break

The magnetic field surrounding PWNe remains largely unconstrained due to the absence of TeV halo counterparts in X-ray and radio observations. Attempts to detect the X-ray halo using eROSITA data, for example, set an upper limit of 4~$\mu$G for the magnetic field in the ISM surrounding B0454+14 due to non-detection. The magnetic field upper limits for other PWNe with TeV halos are estimated to be 1.4--3.1~$\mu$G, suggesting relatively low field strengths in these environments \citep{Khokhriakova2024}. To assess the impact of magnetic field variations, we explore field strengths of 3, 5, and 10~$\mu$G on the non-thermal emission and positron flux. The diffusion coefficient value for different magnetic field strength is shown in Table \ref{table: diffusion coefficient}.

\hfill\break
On top of the standard diffusion of the pairs in propagation, we include the re-acceleration due to energy transfer by scattering on the Alfv\'en wave of the ISM plasma. We use a velocity of the Alfv\'en wave $V_{A}=17$ km/s, as inferred from the CR isotope measurement of the primary to secondary species \citep{Boschini_2017}. However, we expect the effect of the re-acceleration to be negligible.
\hfill\break
\begin{table}
\centering
\begin{tabular}{|c|c|c|c|}
\hline
Magnetic field                                                & 3 $\mu$G & 5 $\mu$G & 10 $\mu$G \\ \hline
Diffusion Coefficient $D_{100\rm{TeV}}$(10$^{27}$cm$^{2}$/s) & 4.1802   & 7.3490   & 22.2027    \\ \hline
Suppression ratio ($D_{\rm{SDZ}}$/$D_{\rm{ISM}}$)             & 0.0027   & 0.0047   & 0.0143    \\ \hline
\end{tabular}
\caption{Diffusion coefficient used for magnetic field strength 3, 5, and 10~$\mu$ G.}
\label{table: diffusion coefficient}
\end{table} 
\hfill\break
Recent research on the self-confinement mechanisms that result from the propagation of charged particles around PWNe suggests that the suppression of diffusion extends from 50 to 100 pc, according to theoretical studies by  \citet{Malkov_2013, Nava_2016, Evoli_2018, Schroer_2022}. There may be a link between the Monogem pulsar and the magnetic disturbance it produces in the surrounding interstellar plasma. Due to this, the size and location of the slow-diffusion bubble may change over time. To account for the expansion of the bubble over the lifetime of the PWN, we estimate that the radius of the bubble grows as the diffusion of the charged particles. Therefore, $r_{1,2}=\mu\sqrt{t}$, where $\mu$ is a constant. To align with the observations made by HAWC, the current values of $r_1$ and $r_2$ must be equal to or larger than the diffusion length of $\sim 100$~TeV electrons/positrons.  

\hfill\break
Motivated by scenarios involving confinement due to the turbulence of a supernova remnant, we explore the possibility that the Monogem Ring defines the SDZ. In this case, the center of the SDZ stays static at the birth location of the Monogem pulsar. We assume the same expansion history of the slow-diffusion bubble. The value of $\mu$ is chosen to assure the current-day SDZ size is close to approximately 70~pc in radius, consistent with current size estimates of the expanding supernova remnant's shell \citep{Thorsett_2003}. 

\hfill\break
In order to cover different scenarios, we have set up the following simulation configurations:
A) In the first configuration, both the pulsar and the SDZ remain stationary and are centered at the current location of B0656+14. The size of the SDZ remains unchanged over time.
B) In the second configuration, both the pulsar and SDZ move with the current proper motion of the pulsar and the size of the SDZ evolves over time.
C) In the third configuration, the PWN moves at the pulsar proper motion. However, the SDZ center remains static at the birthplace of the pulsar. The size of the SDZ expands over time and has a radius of 70~pc at the current time. It is clear that the stationary scenario is not a realistic representation of the actual system. However, with such a setup, it is possible to for us to identify the effect of the size of the SDZ and the relative motion of the PWN and SDZ on the non-thermal emission.
%\OM{This reads well.}

\section{Results}\label{sec:results}
In this section, we explore model parameters by comparing the IC emission with HAWC and Fermi-LAT observations from the Monogem pulsar region. We also discuss how these parameters influence the IC and Synchrotron emission morphology. Our base propagation parameter setup and slow-diffusion-zone model parameters are presented in Table \ref{tab:Pulsar_Model_parameters} and Table \ref{table: diffusion coefficient}. To estimate the pulsar's injection efficiency ($\eta$) for each model, we fit the surface brightness profile in the 8-40~TeV range to the HAWC observation using a traditional $\chi^2$ analysis.
\hfill\break
\subsection{Non-thermal Radiation}%Electromagnetic Emission}
\subsubsection{Senario A: Stationary PWN and Slow-diffusion Bubble}
\begin{figure}
    \centering
    \begin{subfigure}[b]{0.5\textwidth}
        \centering
        \includegraphics[height=0.28\textheight]{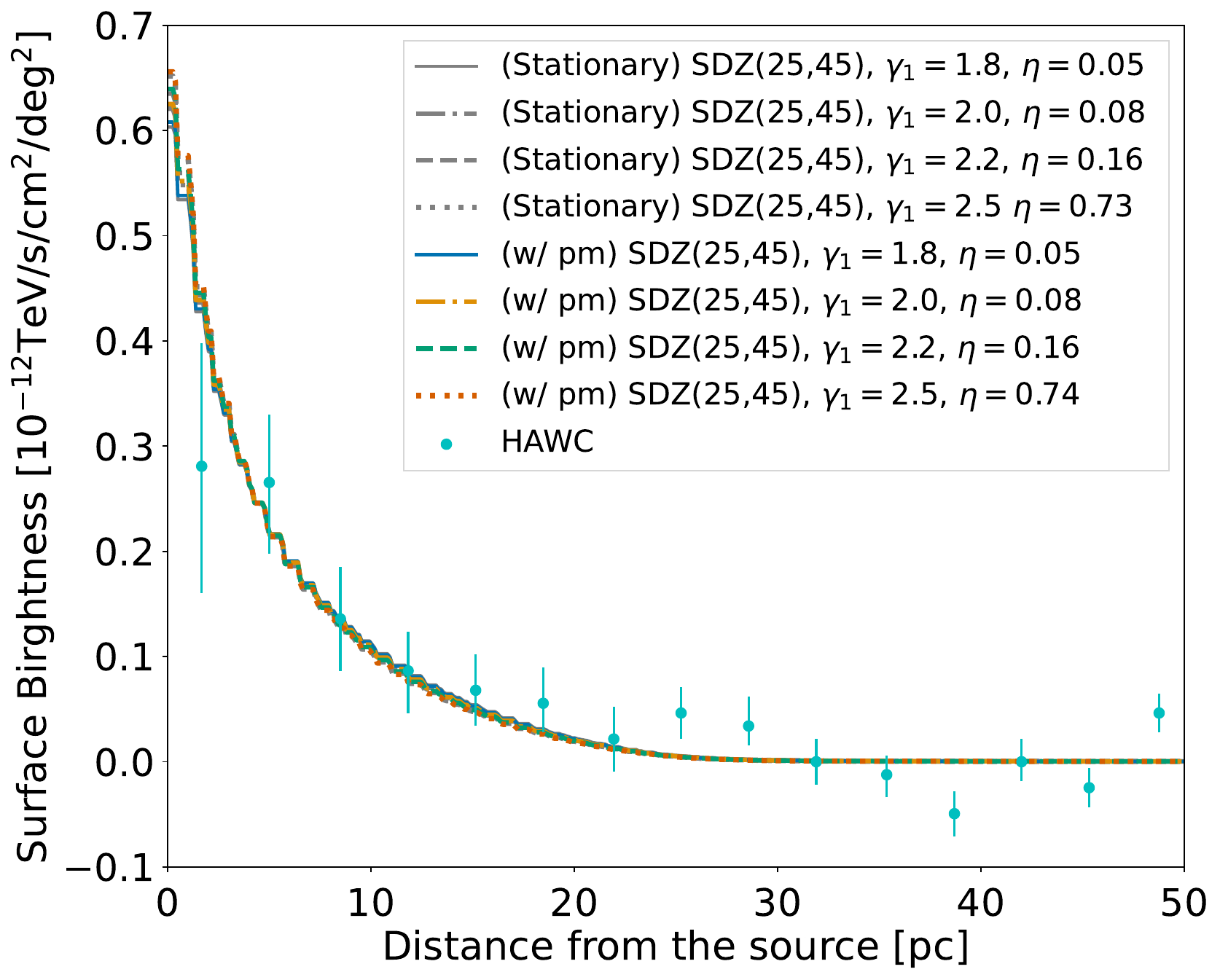}
     \end{subfigure}
     \hfill
     \begin{subfigure}[b]{0.5\textwidth}
         \centering
         \includegraphics[height=0.28\textheight]{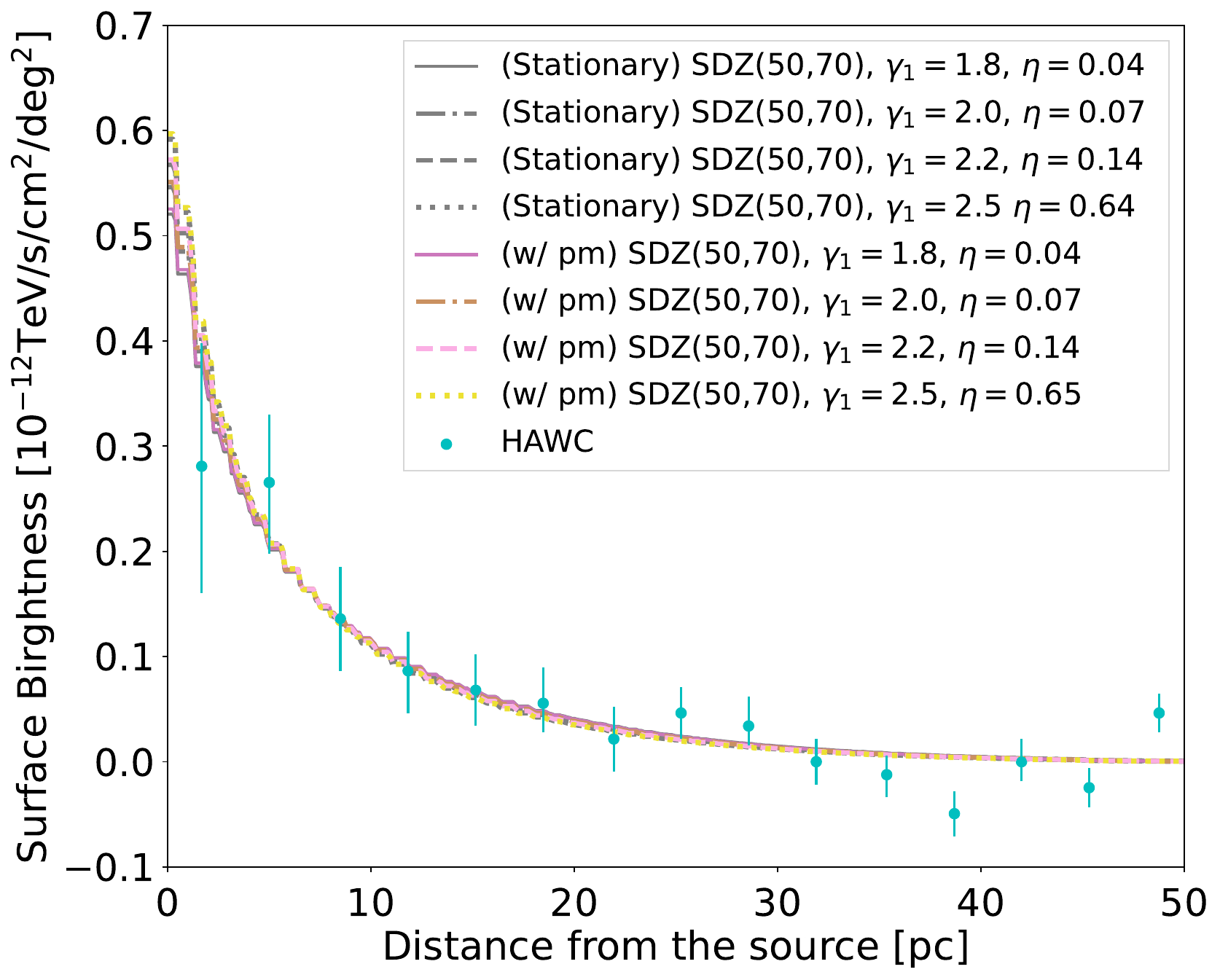}       
     \end{subfigure}
\caption{Surface brightness of IC emission in 8--40~TeV range for stationary scenario (in gray lines) and scenario B where the diffusion bubble moves with the pulsar (in colored lines). The initial time-scale of pulsar $\tau_{0}= 0.1 \tau_{\rm c}$. The magnetic field strength $B= 3~\mu$G. The escape efficiency $\eta$ is tuned to fit blue data points for HAWC observation. \textbf{Top}: SDZ size of (25, 45)~pc. \textbf{Bottom}: SDZ size of (50, 70)~pc. }
     \label{fig: stationary IC}
\end{figure}
We assume that the Monogem pulsar and its slow-diffusion bubble are stationary and located at the Monogem pulsar's present-day position. To construct non-thermal emission maps, we ran various GALPROP simulations using two different sizes of the SDZ: $(r_{1},r_{2}) = (25, 45)$~pc and $(r_{1},r_{2}) = (50, 70)$~pc, as well as four high-energy injection indices: $\gamma_{1}= 1.8, 2.0, 2.2, 2.5$. We created radial profiles from the resulting gamma-ray maps and used them to fit the HAWC data points in the energy range of 8--40~TeV. Our analysis, presented in Figure \ref{fig: stationary IC}, reveals no noticeable variation in the surface brightness with the selected sizes of the SDZ.

\hfill\break
As depicted in Figure \ref{fig: stationary IC}, the required escape efficiency to fit the TeV-halo observation does not differ significantly for SDZ sizes of 25--45~pc and 50--70~pc. We have found that for values of $\gamma_{1}$ between 1.8 to 2.5, a fraction of approximately 5\% to 75\% of the current pulsar spin-down energy must be carried out by the electrons/positrons that escape from the PWN into the ISM to explain the observations made by HAWC. A higher escape efficiency is required for a softer injection spectrum.

\begin{figure}
    \centering
    \begin{subfigure}[b]{0.5\textwidth}
        \centering
        \includegraphics[height=0.28\textheight]{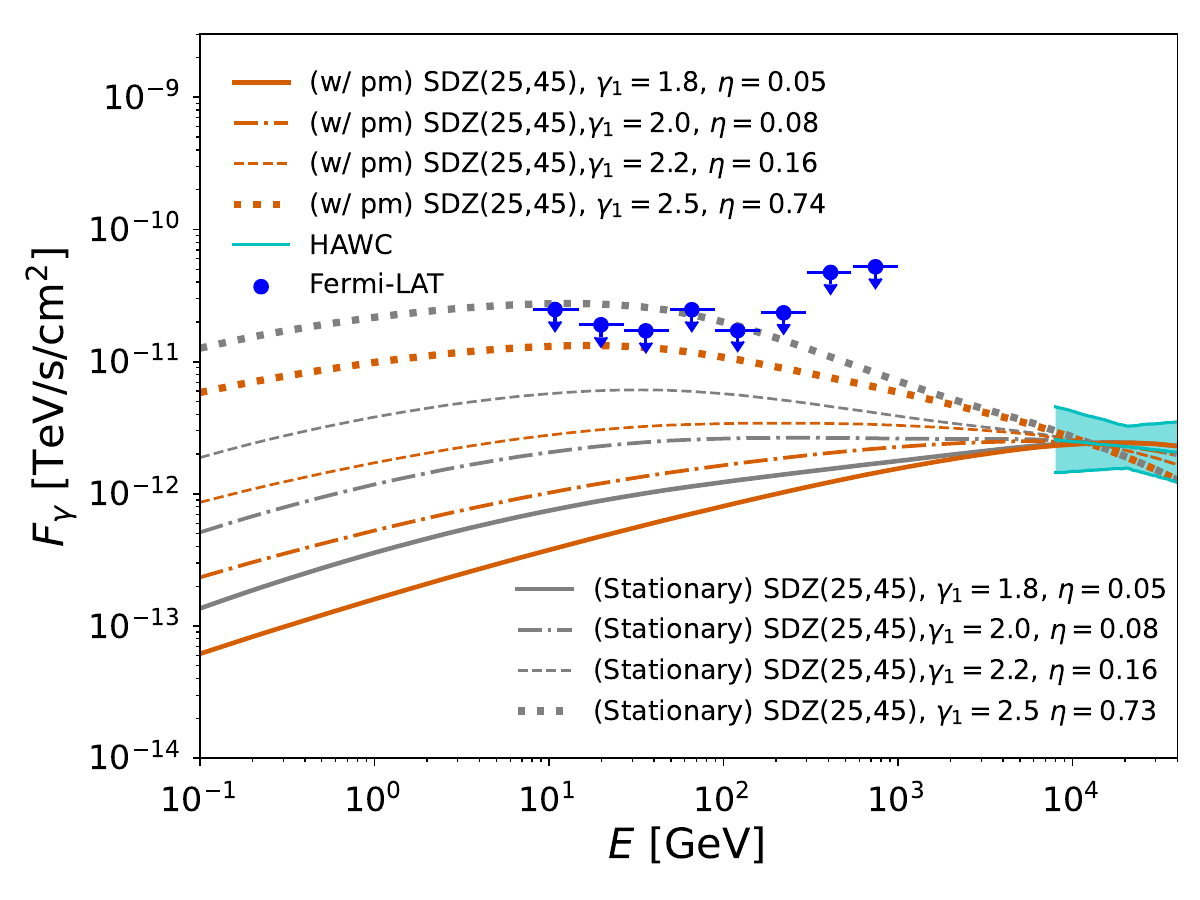}
     \end{subfigure}
     \hfill
     \begin{subfigure}[b]{0.5\textwidth}
         \centering
         \includegraphics[height=0.28\textheight]{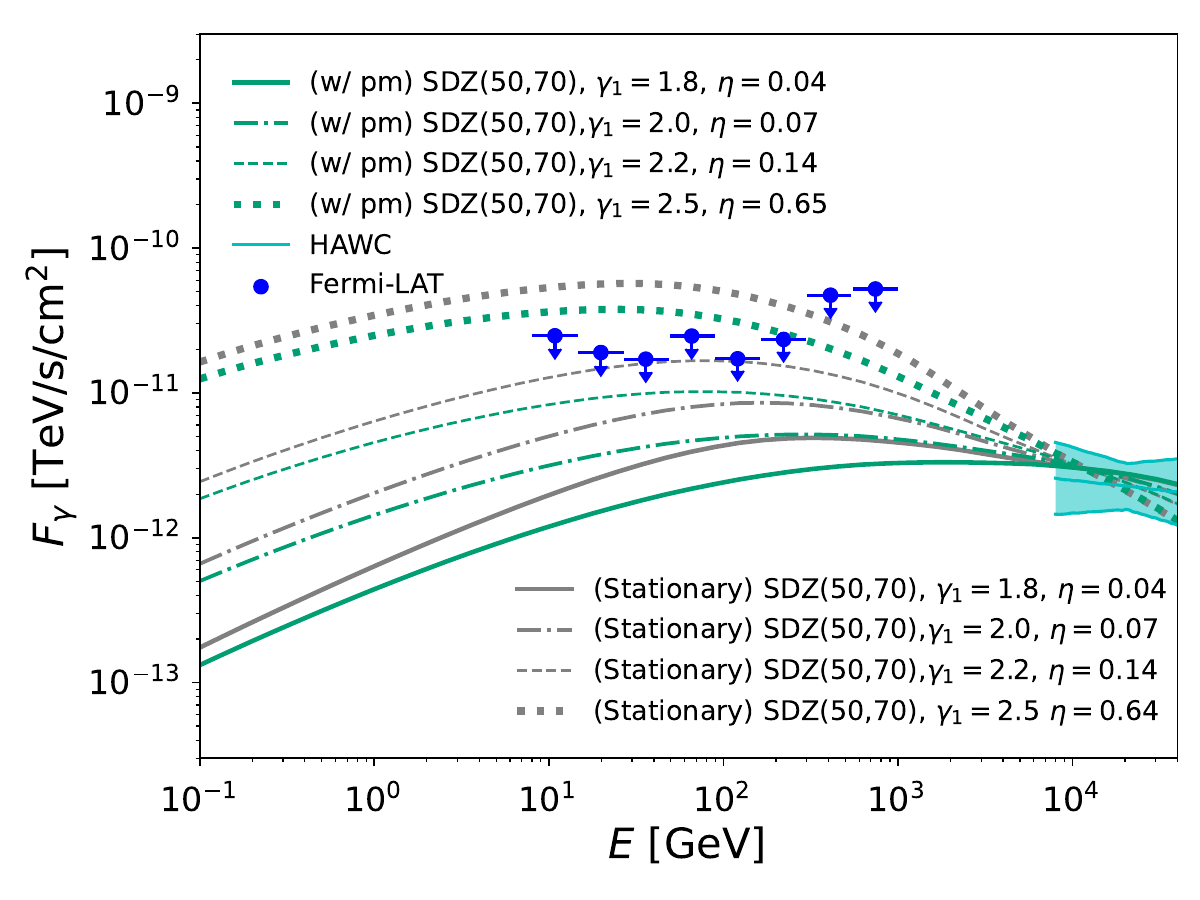}      
     \end{subfigure}
\caption{IC spectra of $\gamma_{1}=$1.8, 2.0, 2.2 and 2.5 of the stationary scenario (in gray lines) and scenario B where the diffusion bubble moves with the pulsar (in colored lines), compared with Fermi-LAT upper-limits \citep{Di_Mauro_2019} and HAWC observation \citep{HAWC_2017}. The initial time-scale of pulsar $\tau_{0}= 0.1 \tau_{\rm c}$. The magnetic field strength $B= 3~\mu$G. \textbf{Top}: SDZ size of (25, 45)~pc. \textbf{Bottom}: SDZ size of (50, 70)~pc. }

\label{fig:IC_spectra_stationary_pm}
\end{figure}

\hfill\break
Figure~\ref{fig:IC_spectra_stationary_pm} shows the energy spectra of IC emissions in the energy range of 10~GeV to 25~TeV from the $10^\circ$ region around the pulsar, compared with the HAWC observation in the range of 8--40~TeV and the Fermi-LAT upper limits derived by \citet{Di_Mauro_2019}. For each $\gamma_{1}$, the IC flux level of SDZ size (50,70)~pc is higher than that of (25, 45)~pc in GeV energies. A softer injection spectrum also yields a higher flux in the GeV--TeV energies. The case when $\gamma_{1}=2.5$, and SDZ size of (50, 70)~pc is ruled out due to the Fermi-LAT upper limits. For the smaller size SDZ, while the stationary case of $\gamma_{1}=2.5$ went higher than the Fermi-LAT upper-limit, the flux remained consistent once the proper motion of the pulsar is included, as we discuss below, thus is still possible. Any softer injection spectrum than $\gamma_{1}=2.5$ is disfavored to fit with the current observations and constraints. 

\begin{figure}
    \centering
    \begin{subfigure}[b]{0.5\textwidth}
        \centering
         \includegraphics[height=0.29\textheight]{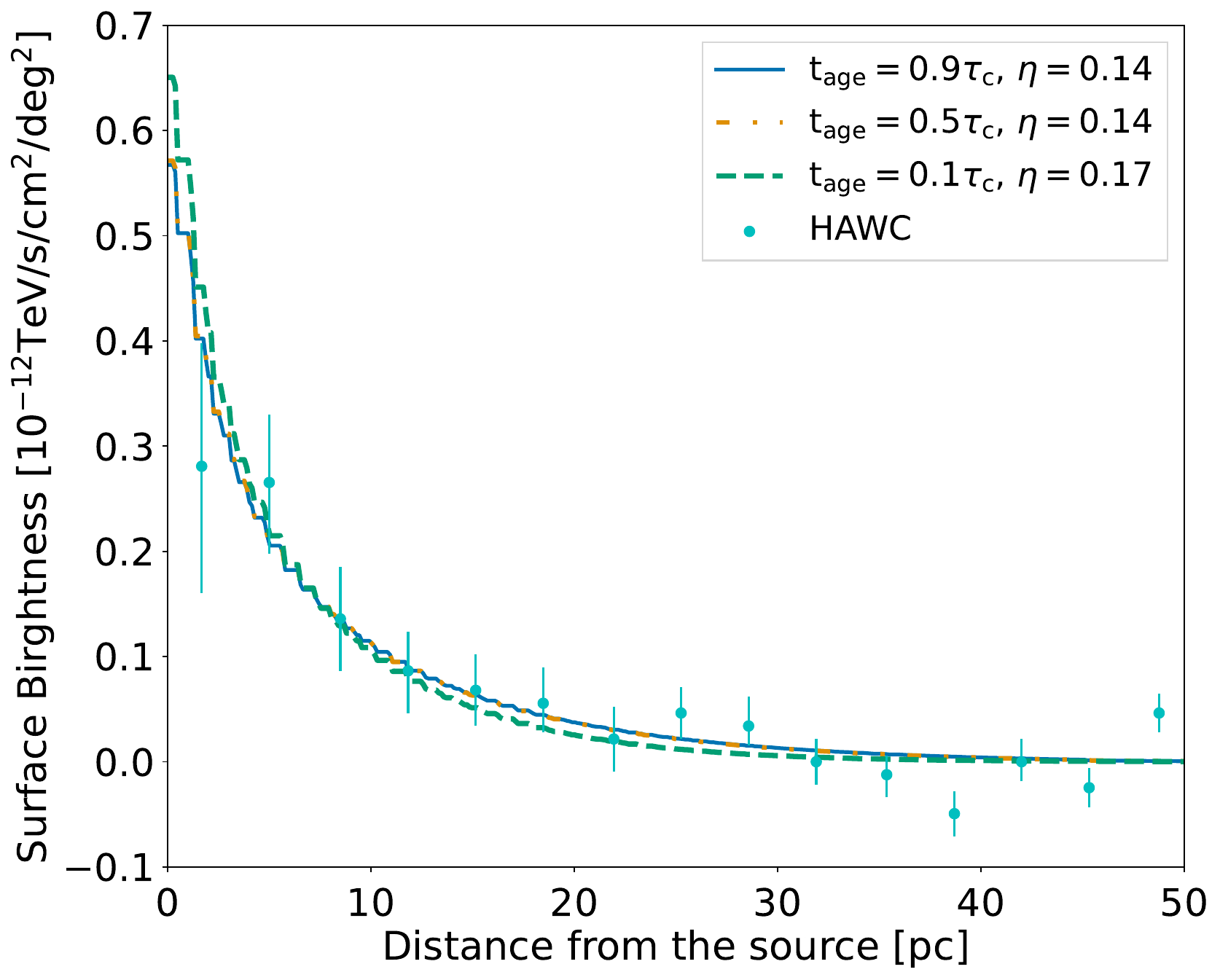}
%        \subcaption{}
     \end{subfigure}
     \hfill
     \begin{subfigure}[b]{0.50\textwidth}
         \centering
          \includegraphics[height=0.28\textheight]{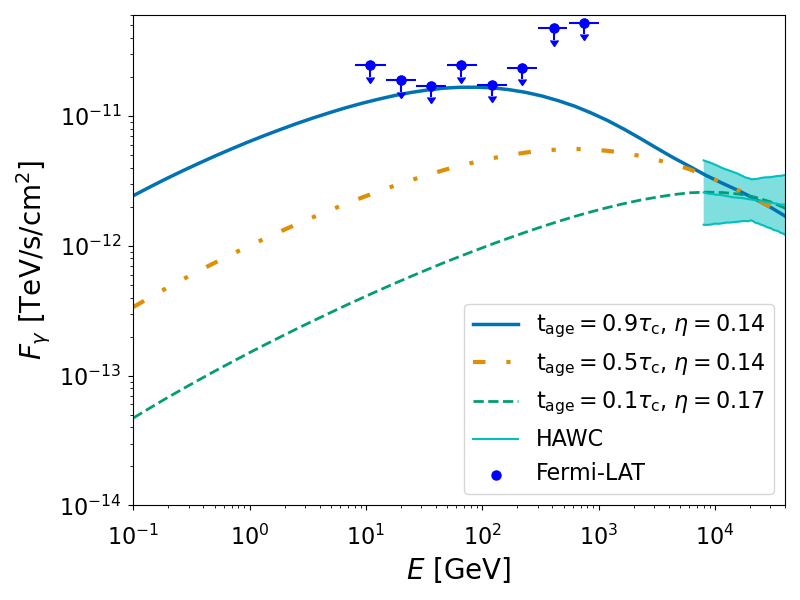}
%         \subcaption{}
     \end{subfigure}
\caption{IC emission of varied $\tau_{0}$ of stationary pulsar and SDZ, compared with HAWC observation \citep{HAWC_2017} and Fermi-LAT upper-limits \citep{Di_Mauro_2019}. The size of SDZ is (50,70) pc. The magnetic field strength is B= 3 $\mu$G. The injection index $\gamma_{1}=2.2$. \textbf{Top:} Surface brightness in 8- 40 TeV for $t_{\rm age} =$0.9, 0.5, 0.1 $\tau_{\rm c}$ fitted to HAWC 
observation. \textbf{Bottom:} IC spectra of $t_{\rm age} =$ 0.9, 0.5, 0.1 $\tau_{\rm c}$ compared to HAWC observation and Fermi-LAT upper-limit.
}
\label{fig:IC_5070_age_varied}
\end{figure}

\hfill\break
We present the IC emission for Monogem pulsar assuming varying true ages. We adjusted the parameters $t_{\rm age}$ and $\tau_{0}$ based on Eq. \ref{tau_0} to simulate their emission. Specifically, we compared the emissions from $\tau_{0}$ values of 0.1, 0.5, and 0.9 $\tau_{\rm c}$, which correspond to $t_{\rm age}$ values of 99, 55, and 11 kyr, respectively. In this simulation, the pulsar is assumed to be stationary. The simulation results are shown in Figure \ref{fig:IC_5070_age_varied}, which displays the IC surface brightness and spectra. In the energy range of 8--40 TeV, the surface brightness are almost identical for $t_{\rm age} = 0.9, 0.5$ and $0.1 \tau_{\rm c}$. The cooling time of $e^{\pm}$ at 100 TeV is $\sim 10$~kyr for a magnetic field of $B = 3~\mu\text{G}$ based on Eq. \ref{eq:cooling_time}, which is shorter than the injection timescale of the particles for $t_{\rm age} = 0.9, 0.5 \tau_{\rm c}$, and is comparable for $t_{\rm age} = 0.1 \tau_{\rm c}$. Thus, the surface brightness of the HAWC observation can be reproduced in all three cases. However, the energy spectrum of the $10^{\circ}$ region around the PWN shows an enhancement in the sub-GeV--GeV range as the pulsar true age increases.
\subsubsection{Scenario B: Moving PWN and Slow-diffusion Bubble}
In this section, we examine the effect of the Monogem pulsar's proper motion on the anticipated spectra and spatial morphology of the non-thermal radiation. Furthermore, we investigate the possibility of detecting the source of the slow-diffusion bubble around the PWN in the non-thermal emission.

\hfill\break
We begin by discussing the scenario in which the propagation of pairs injected by Monogem self-induces the slow diffusion bubble. In this context, the SDZ will always be centered at the pulsar location and will move with the same proper motion. Figure \ref{fig: stationary IC} and Figure \ref{fig:IC_spectra_stationary_pm} compare the surface profile and the IC energy spectra of the stationary scenario and scenario B, assuming different injection spectral indices. The surface brightness of the halo, after including the proper motion of the pulsar and the evolution of the diffusion bubble, is identical to the stationary case. This is likely due to the small transverse proper motion of the Monogem pulsar. The injection efficiencies of the two scenarios are also mostly identical. However, the energy spectra exhibit lower flux compared to the stationary case because the evolution of the SDZ is taken into account. A smaller SDZ in the past led to greater escape of the accelerated $e^{\pm}$ from the bubble, thereby lowering the IC emission flux around the PWN. An injection index of $\gamma_{1} = 2.5$ is ruled out for an SDZ size of (50, 70)~pc due to overshooting the Fermi-LAT upper limit. However, for a smaller SDZ size of (25, 45)~pc, this index is still viable, with a fairly high injection efficiency of $\eta = 0.74$.

\begin{figure*}
\centering
\begin{subfigure}[t]{.49\textwidth}
  \centering  \includegraphics[width=.8\linewidth]{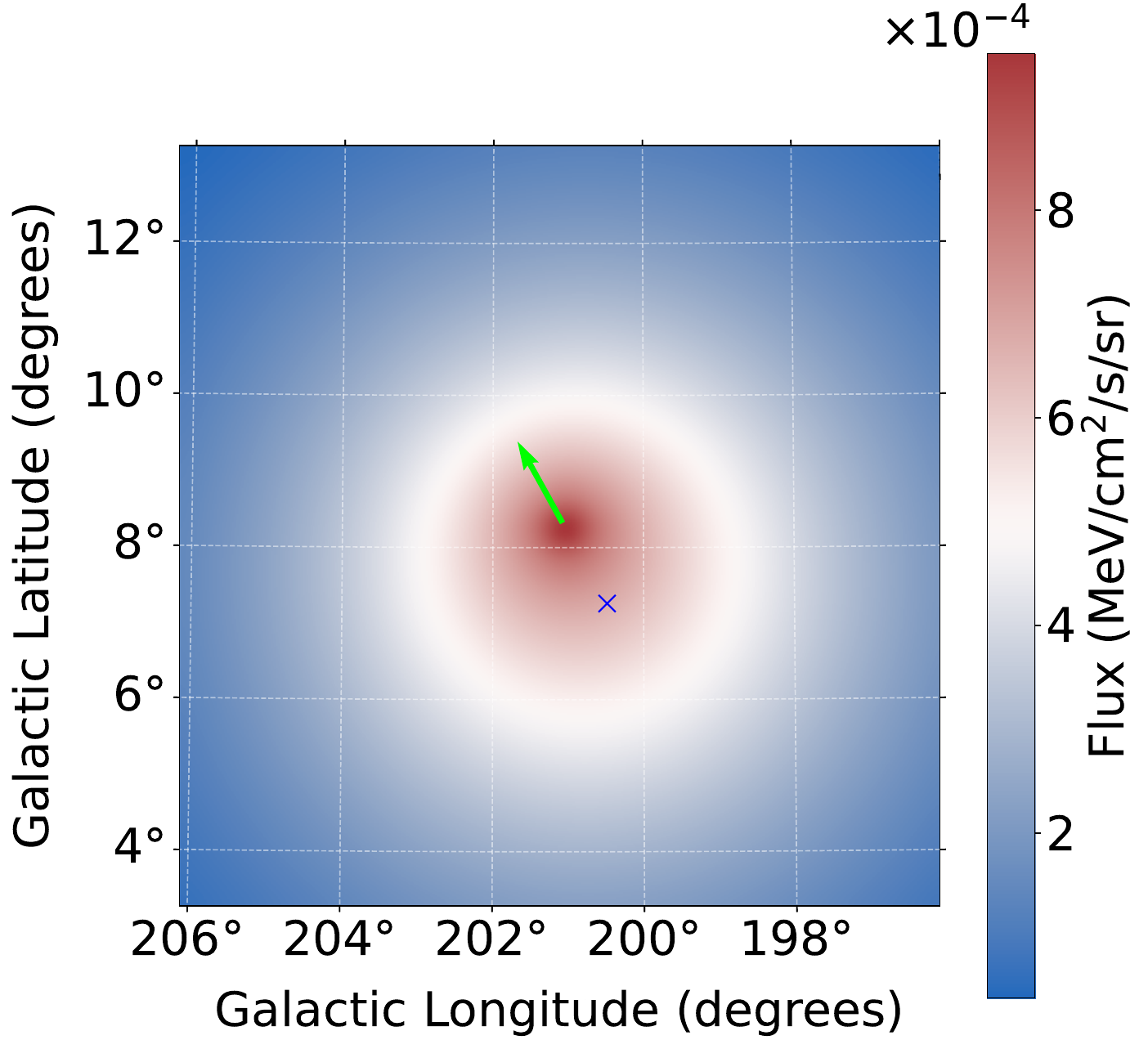}
  \caption{IC emission at 10 GeV}
  \label{fig:IC_pm_10GeV}
\end{subfigure}%
\begin{subfigure}[t]{.49\textwidth}
  \centering  \includegraphics[width=.8\linewidth]{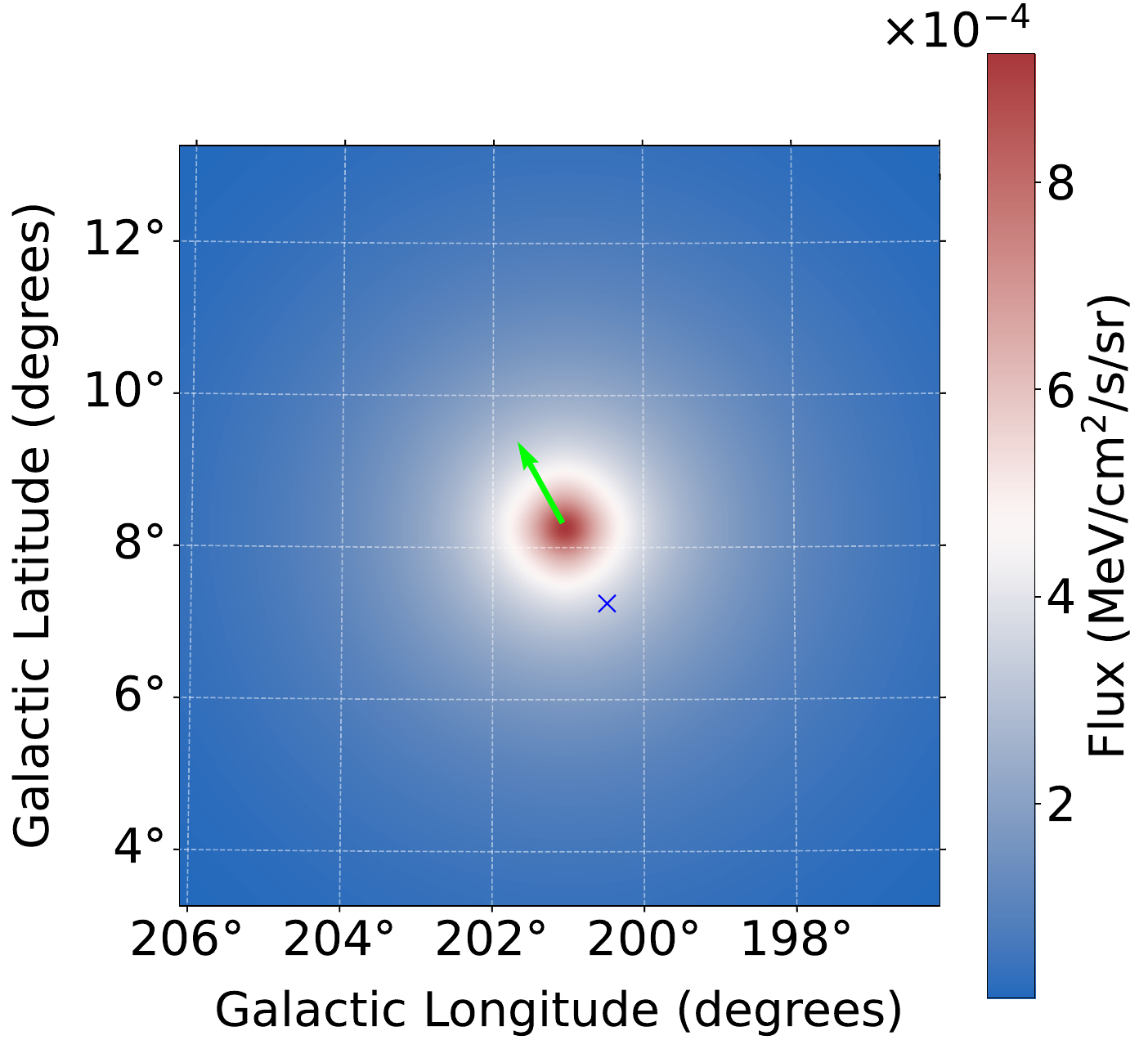}
  \caption{IC emission at 10 TeV}
  \label{fig:IC_pm_10TeV}
\end{subfigure}
\begin{subfigure}{.49\textwidth}
  \centering  \includegraphics[width=.8\linewidth]{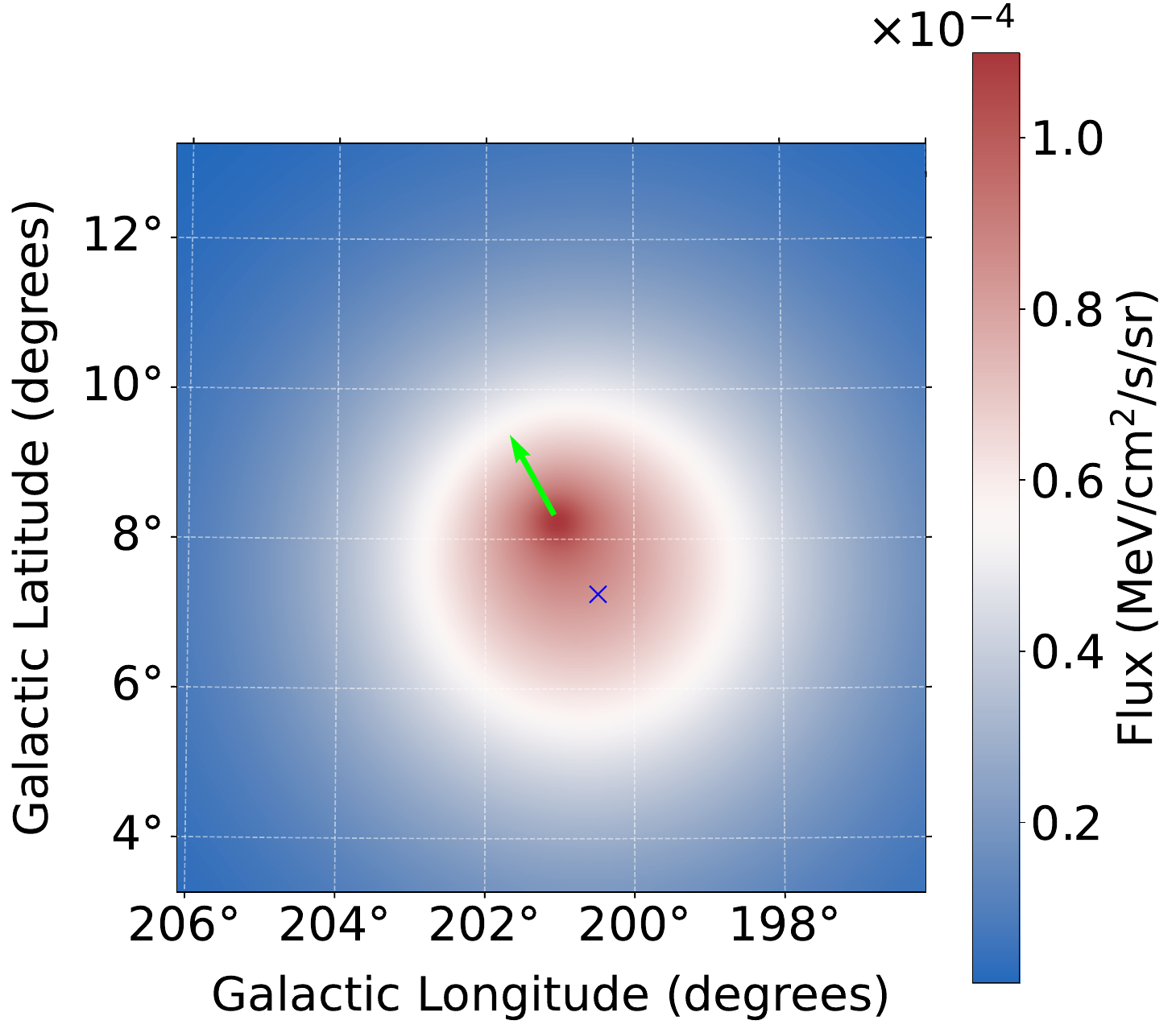}
  \caption{Synchrotron emission at 100 GHz}
  \label{fig:sync_pm_100GHz}
\end{subfigure}%
\begin{subfigure}{.49\textwidth}
  \centering  \includegraphics[width=.8\linewidth]{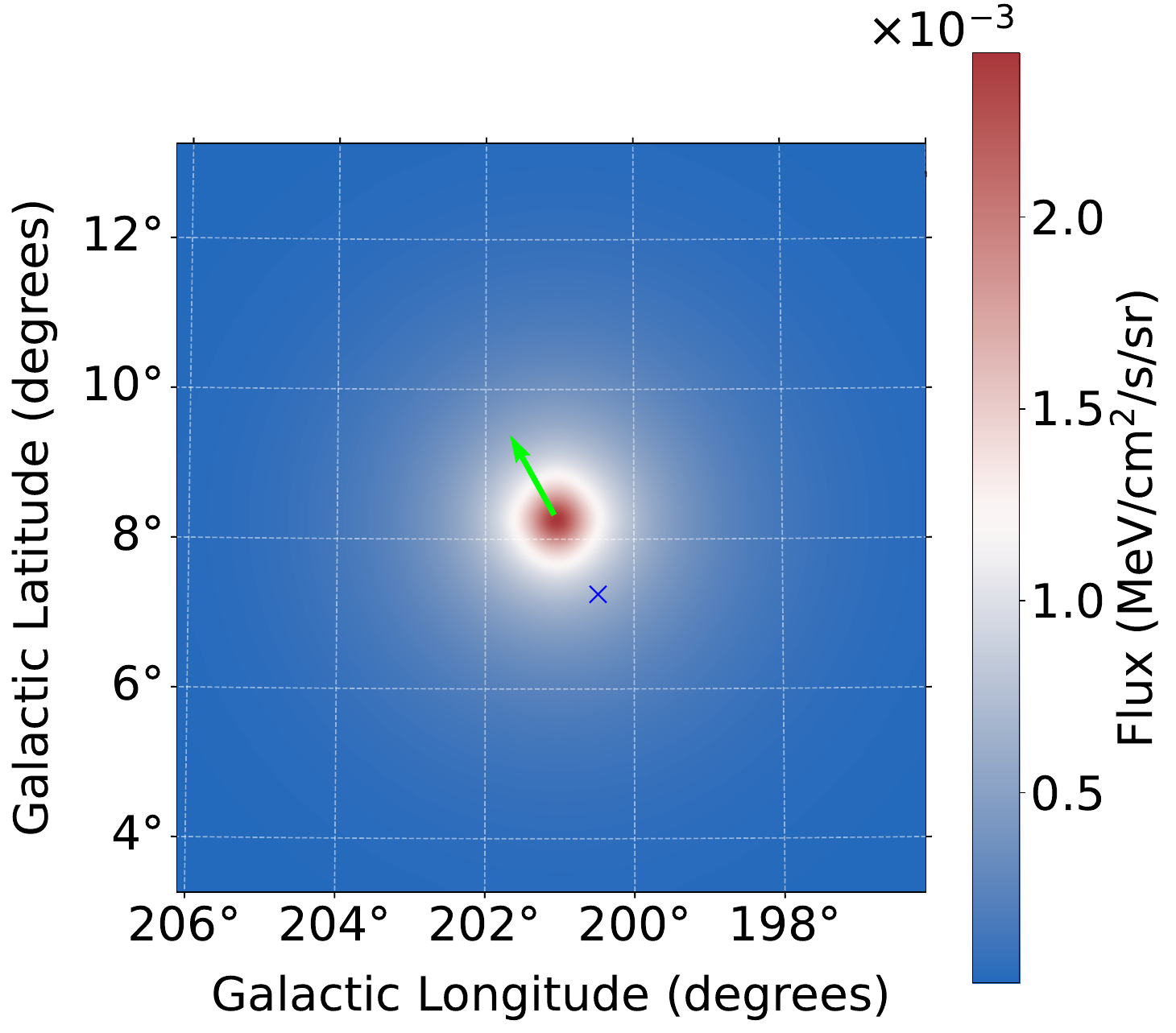}
  \caption{Synchrotron emission at 5 keV}
  \label{fig:sync_emission_5keV}
\end{subfigure}\\
\caption{Intensity maps at different wavelengths for the region around Monogem pulsar taking into account the pulsar proper motion. The center of the SDZ is always at the pulsar (Scenario B). The SDZ has a size of $(r_{1},r_{2})=(50,70)$~pc at current time. The pulsar true age is $t_{\rm age}= 0.9\tau_{\rm c}$. The magnetic field strength is $B=3~\mu$G, and injection index $\gamma_{1}=2.2$. The lime vector shows the pulsar proper motion, and the blue cross marks the estimated pulsar birthplace. Asymmetry in the emissions due to the proper motion of the PWN and SDZ is significant for 10~GeV IC emission (a), and 100~GHz synchrotron emission (c). While higher symmetry is found in IC emission at 10~TeV (b), and synchrotron emission at 5~keV (d).
}
\label{fig: pm_emission_map}
\end{figure*}

\hfill\break
Figure \ref{fig: pm_emission_map} displays the intensity map of IC and synchrotron emissions near the PWN within a radius of approximately $\sim 30$ pc. The extended halo emission is visible in the GeV-TeV, as well as radio and X-ray range owing to synchrotron emission. As can be seen in this figure, the IC emission at 10 GeV and the synchrotron emission at 100 GHz exhibit roughly spherical emission with the centroid offsetting from the current location of the pulsar. This offset is aligned with the pulsar proper motion direction. 

\hfill\break
Higher energy observations, such as the IC mission at 10 TeV (top-right panel) and synchrotron emission at 5 keV (bottom-right panel), show more spherically symmetric profiles at the current pulsar location due to the short cooling time of the TeV electrons/positrons. The TeV-halo around Monogem observed by HAWC has a rather lobbed profile~\citep{HAWC_2017}, suggesting that the asymmetry is not due to the proper motion of the pulsar but possibly caused by a spatial asymmetry in the diffusion coefficient in the ISM, or convection due to a local or Galactic-scale wind. We will discuss the former hypothesis in Section \ref{sec: moving PWN fixed SDZ}. For the latter case, the expected wind velocity in the environment to modify the emission morphology in the TeV range is greater than 1600 km/s, which is unlikely to be achieved.

\begin{figure}
     \centering
     \begin{subfigure}[t]{0.45\textwidth}
         \centering        
         \includegraphics[height=0.28\textheight]{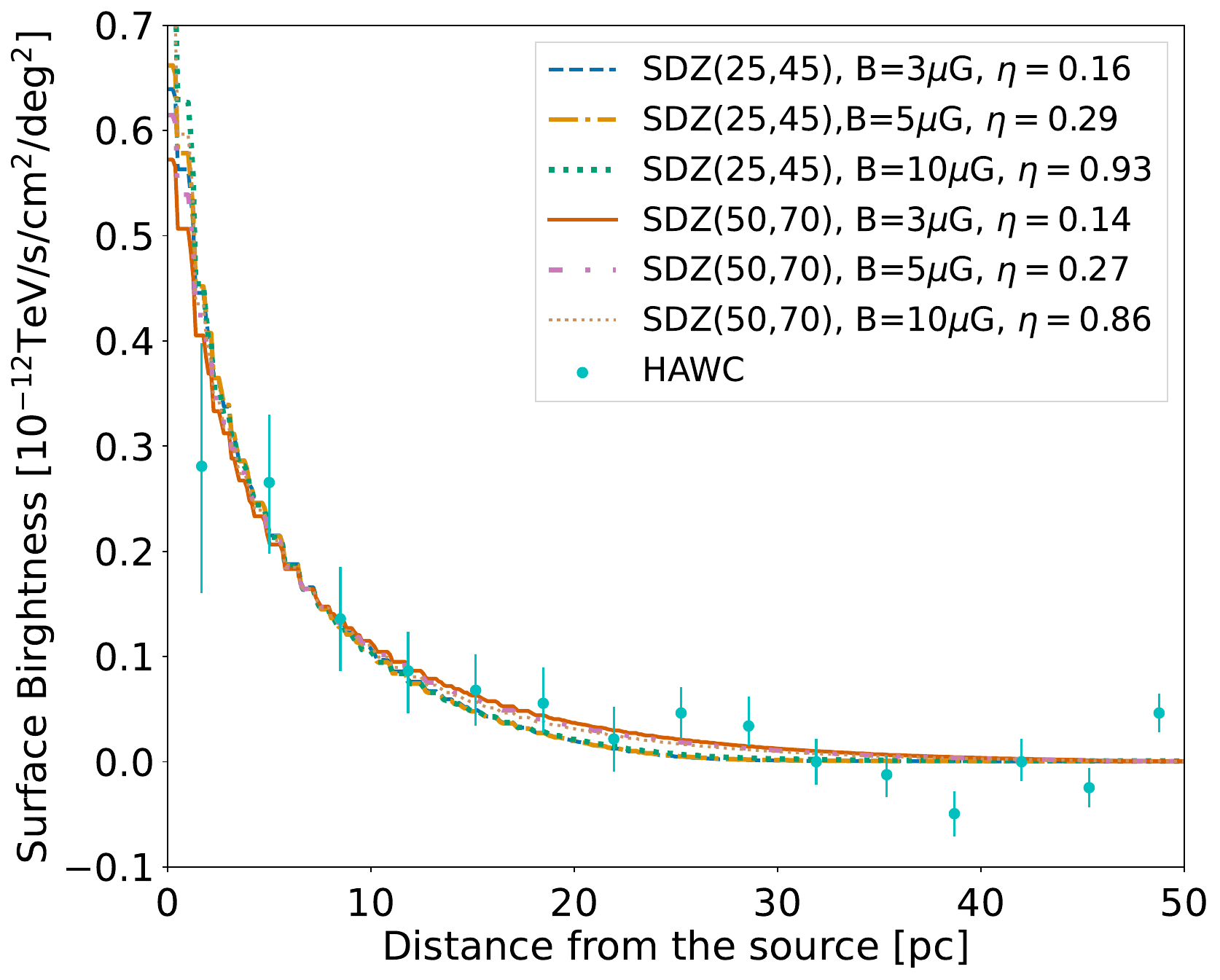}
         \label{fig:IC SB with pm}
     \end{subfigure}%
    \hfill
     \begin{subfigure}[t]{0.45\textwidth}
         \centering
         \includegraphics[height=0.27\textheight]{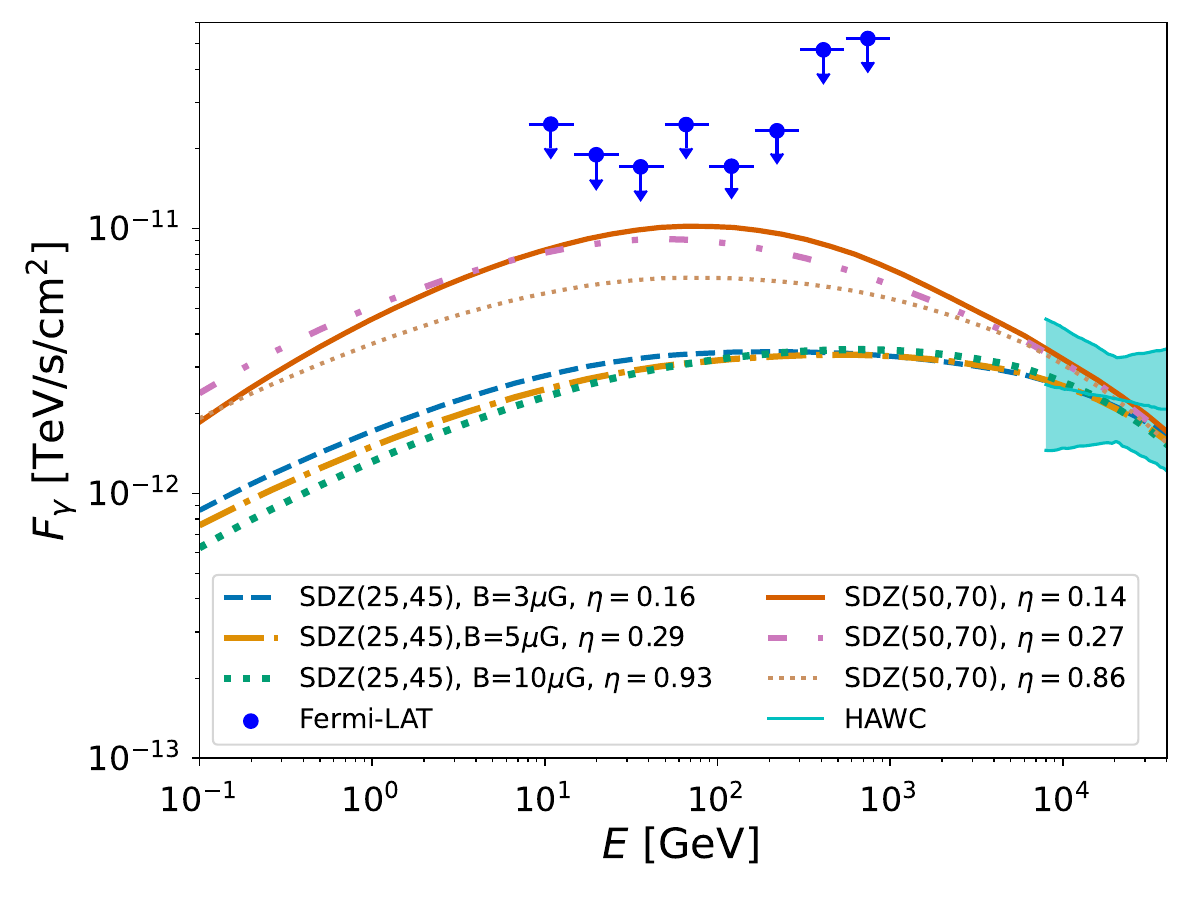}
         \label{fig: IC spec with pm}
     \end{subfigure}\caption{IC emission of scenario B that the slow diffusion bubble moves with the pulsar proper motion. The injection spectra power index $\gamma_{1}=2.2$. The true age of the pulsar $t_{\rm age}= 0.9 \tau_{\rm c}$. We show the expected emission of SDZ sizes of (25,45) pc and (50,70) pc, and magnetic field strength of 3~$\mu$G, 5~$\mu$G, and 10~$\mu$G. \textbf{Top:} Surface brightness in 8-40~TeV in terms of distance from the current location of Monogem pulsar. \textbf{Bottom:} IC spectra of 10$^{\circ}$ region around Monogem pulsar, compared with Fermi-LAT upper-limits \citep{Di_Mauro_2019} and HAWC observation from the region \citep{HAWC_2017}.}
\label{fig: IC_pm_Bfield}
\end{figure}

\begin{figure}
    \centering
    \begin{subfigure}[b]{0.5\textwidth}
        \centering
         \includegraphics[height=0.27\textheight]{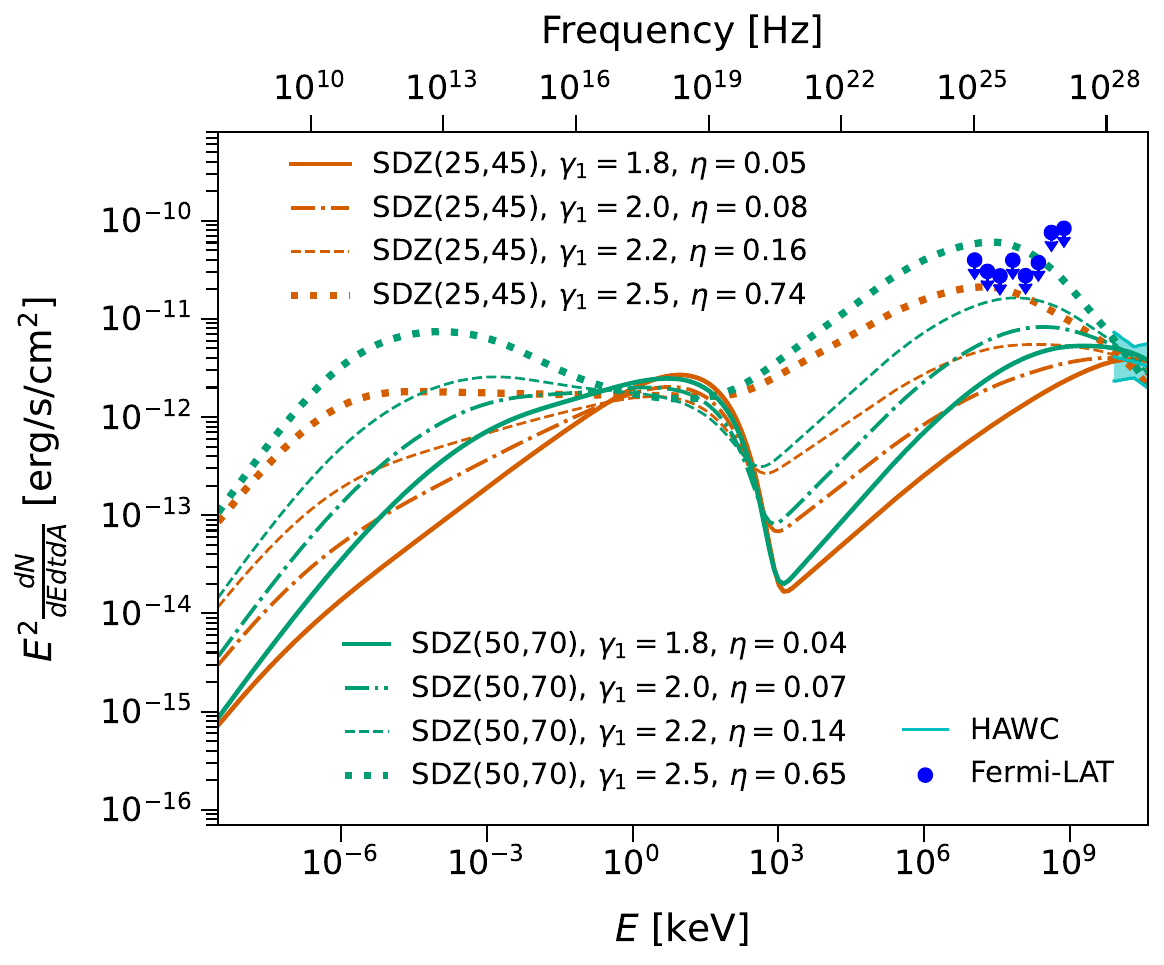}
%        \subcaption{}
     \end{subfigure}
     \hfill
     \begin{subfigure}[b]{0.50\textwidth}
         \centering
          \includegraphics[height=0.27\textheight]{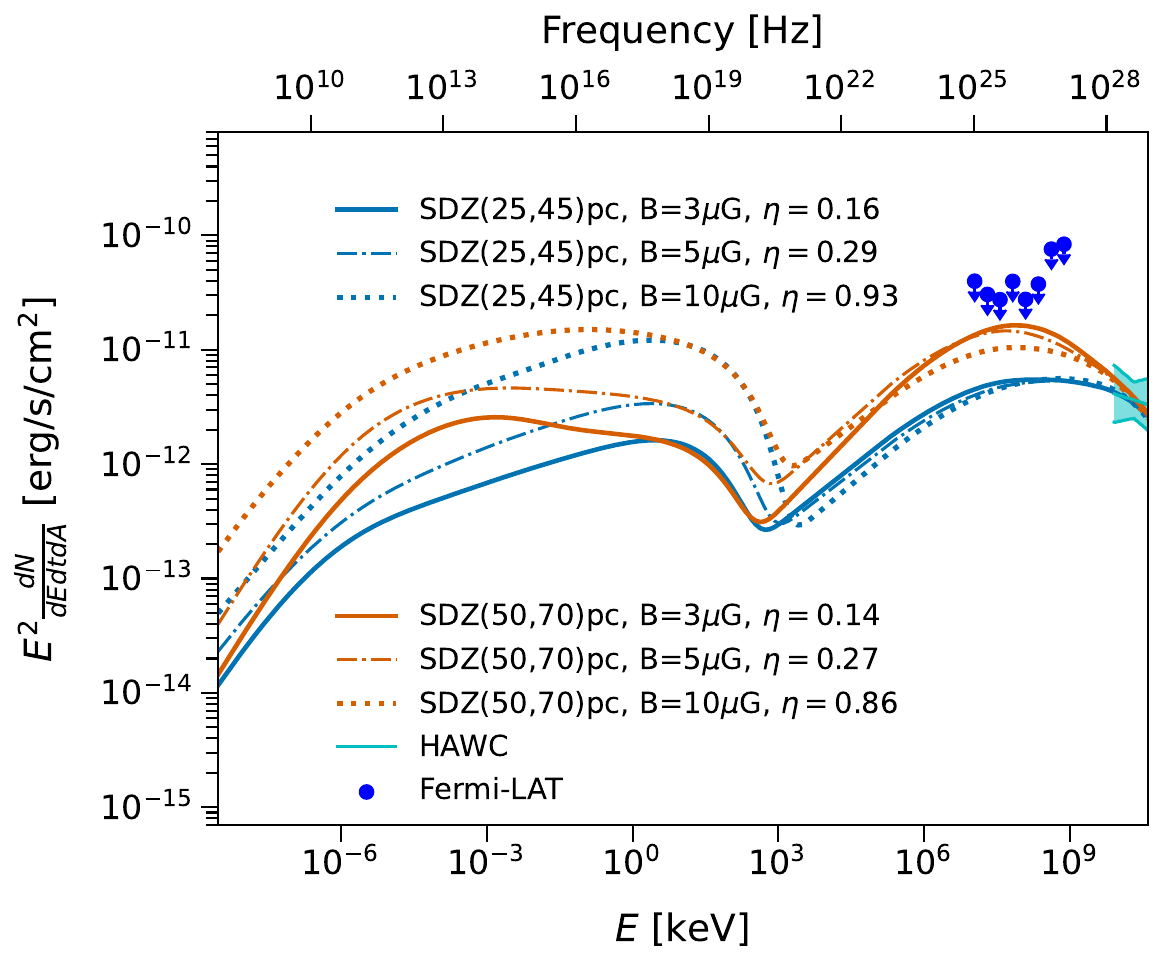}
%         \subcaption{}
     \end{subfigure}
     \hfill
     \begin{subfigure}[b]{0.50\textwidth}
         \centering          \includegraphics[height=0.27\textheight]{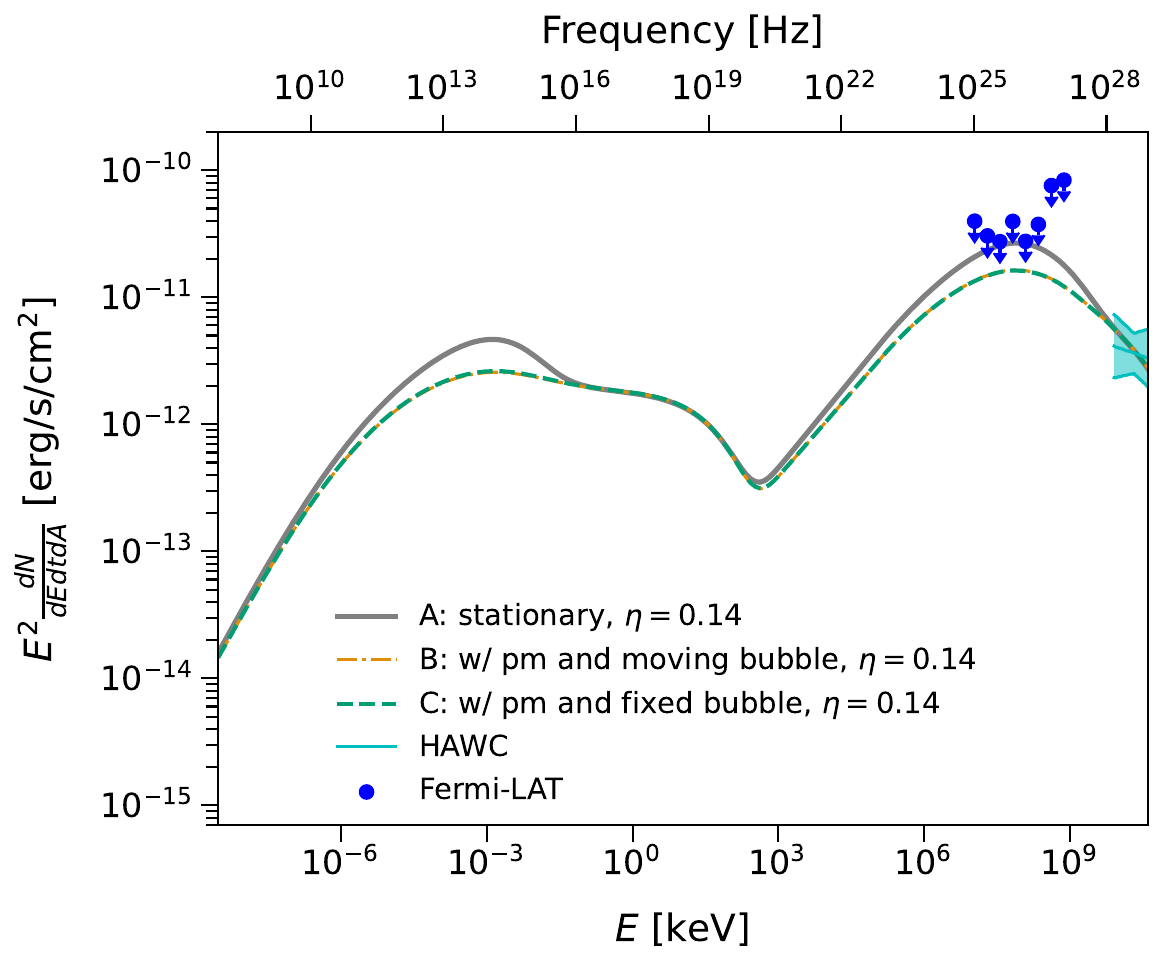}
%         \subcaption{}
     \end{subfigure}
\caption{Synchrotron and IC emission spectra of $10^{\circ}$ region around the PWN  modeled under different parameter combinations. The pulsar's true age is fixed at $t_{\rm age} = 0.9\tau_{\rm c}$ in all cases. \textbf{Top:} Emission spectra for various injection indices, $\gamma_{1} = 1.8$, 2.0, 2.2, and 2.5, in scenario B, with a fixed magnetic field of $B = 3~\mu$G. \textbf{Middle:} Emission spectra for different magnetic field strengths, $B = 3$, 5, and $10~\mu$G, in scenario B, with an injection index $\gamma_{1} = 2.2$. \textbf{Bottom:} Emission spectra for scenarios A (stationary), B, and C. The injection index is $\gamma_{1}=2.2$, and the magnetic field is $B=3~\mu G$.}
\label{fig: sync_IC}
\end{figure}

\hfill\break
We examined how the strength of the magnetic field affects non-thermal emission and positron flux on Earth. Specifically, we tested three random magnetic field strengths (3~$\mu$G, 5~$\mu$G, and 10~$\mu$G) in the solar vicinity, assuming an injection index of $\gamma_{1}=2.2$. A stronger magnetic field reduces the cooling time of electrons/positrons, causing them to lose energy more quickly through synchrotron emission near the PWN. Consequently, the diffusion coefficient needs to be adjusted to account for the TeV halo extension. Based on the estimated diffusion coefficient shown in Table \ref{table: diffusion coefficient}, the surface brightness around the Monogem PWN is reproduced, as shown in the upper panel of Figure \ref{fig: IC_pm_Bfield}. The lower panel shows the IC energy spectrum for different magnetic field strengths with SDZ sizes of (25, 45)~pc and (50, 70)~pc. The flux level differences reflect the size of the SDZ, not the differences in magnetic field strength. The middle panel in Figure \ref{fig: sync_IC} shows the energy spectrum extending to the synchrotron emission from GeV to sub-eV. The level of the synchrotron peak at $\sim 100$~ keV is higher with a stronger magnetic field. 

\hfill\break
The synchrotron spectra predicted for injection index $\gamma_{1}=$ 1.8, 2.0, 2.2, and 2.5 from 10$^{\circ}$ region around the PWN are shown in the top panel of Figure \ref{fig: sync_IC}, together with the corresponding IC emission. Similar to the IC emission, by increasing the size of the SDZ from (25, 45)~pc to (50, 70)~pc, the synchrotron emission flux is enhanced towards lower energy range (sub-eV to sub-keV). The larger SDZ confines lower energy pairs to cool through emission before diffusing away from the SDZ, thus a flatter spectra profile in the eV--keV range is expected. The emission intensity level and spectra feature are similar at keV energies for SDZ sizes of (25, 45) pc, and (50, 70) pc, as well as for different injection indices. If we were going to disentangle the model parameters from observational data, the thermal components in lower than keV energies would make it challenging.
\subsubsection{Scenario C: Moving PWN and Fixed Slow-diffusion Bubble}\label{sec: moving PWN fixed SDZ}
To evaluate the case in which the slow-diffusion zone is related to the expansion of its parent SNR ``the Monogem Ring'', we fixed the SDZ at the Monogem pulsar's birth location and left it static in our simulation. Since we assume constant proper motion of the pulsar and zero radial velocity, the center of the SDZ slightly differs from the estimated center of the Monogem Ring (that comes from observations). Despite this, our simulation should capture the system's evolution well. The SDZ has a current radius of 70~pc, consistent with the radius of the Monogem Ring in X-rays. Assuming an age of 0.9$\tau_{\rm c}$, and no line of sight velocity, we estimate that Monogem traveled $\sim6$~pc from its birthplace, making it well located inside the SDZ, as suggested by observations \citep{Knies_2018}. However, we cannot exclude the possibility that the pulsar has a significant line-of-sight velocity component. Consequently, while the projected position of the pulsar appears within the Monogem Ring, the actual position of the pulsar may lie outside the SNR. 

\begin{figure*}
\centering
\begin{subfigure}[t]{.49\textwidth}
  \centering  \includegraphics[width=.8\linewidth]{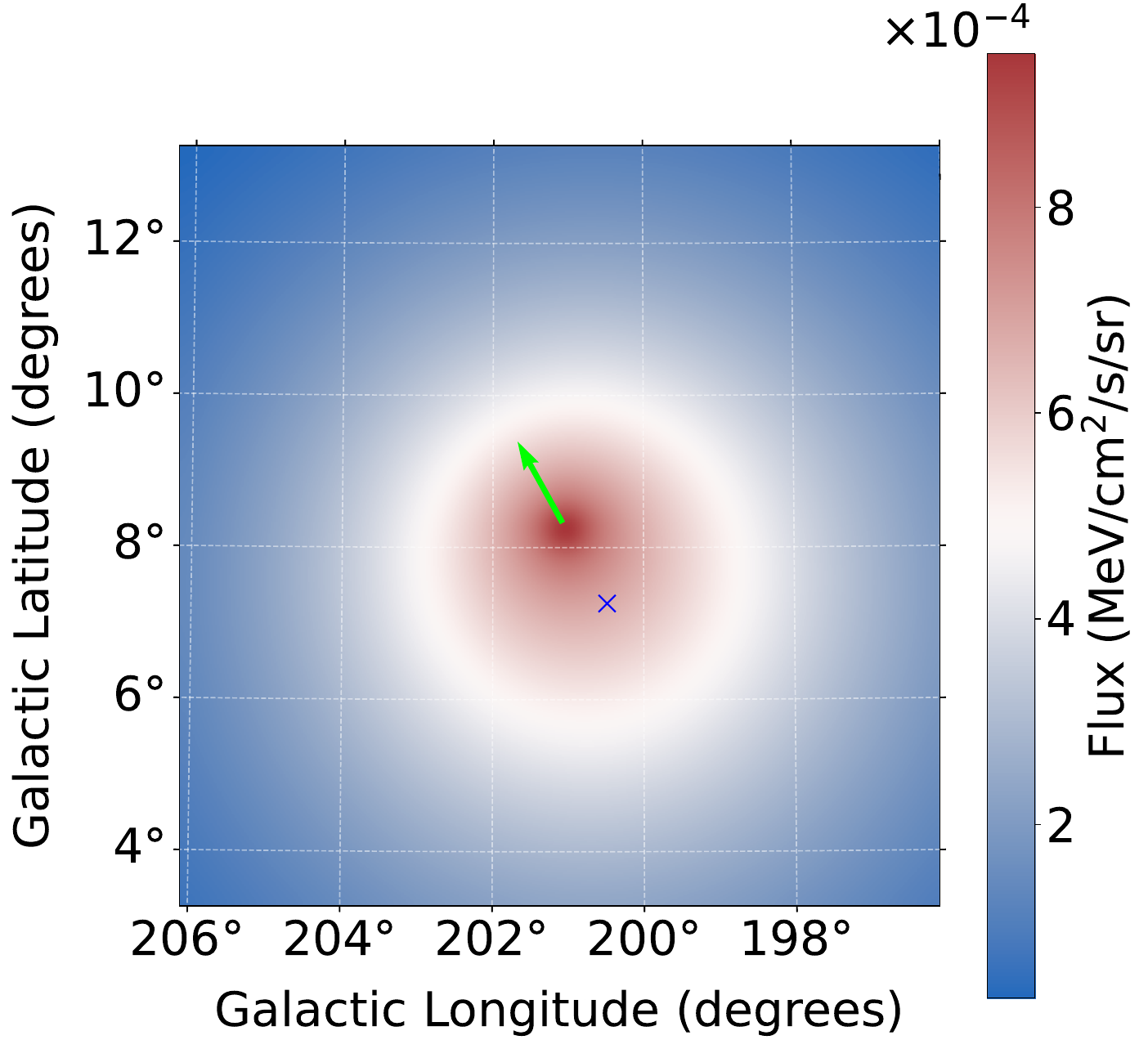}
  \caption{IC emission at 10 GeV}
  \label{fig:IC_pm_fixed_10GeV}
\end{subfigure}%
\begin{subfigure}[t]{.49\textwidth}
  \centering  \includegraphics[width=.8\linewidth]{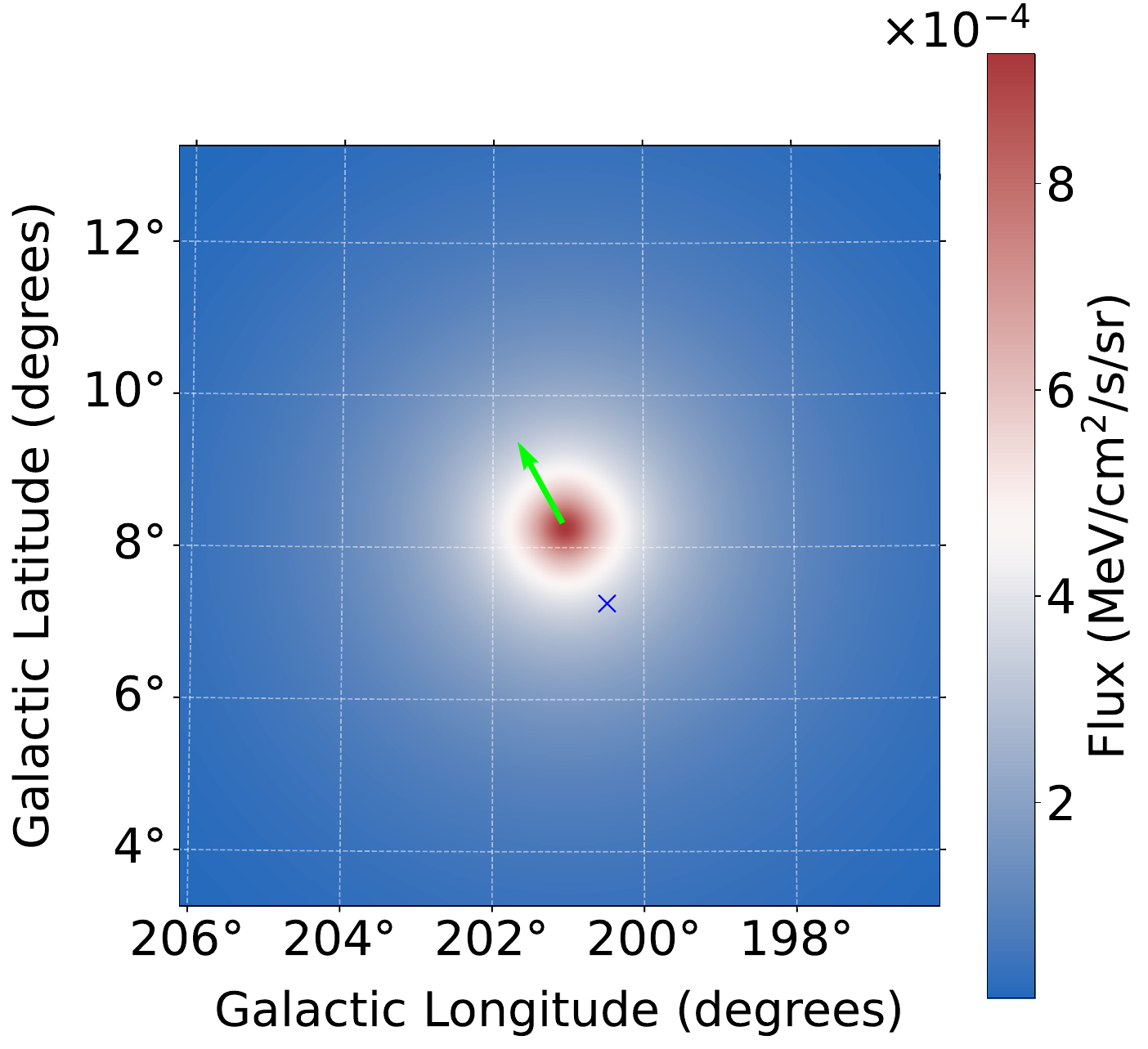}
  \caption{IC emission at 10 TeV}
  \label{fig:IC_pm_fixed_10TeV}
\end{subfigure}
\begin{subfigure}{.49\textwidth}
  \centering  \includegraphics[width=.8\linewidth]{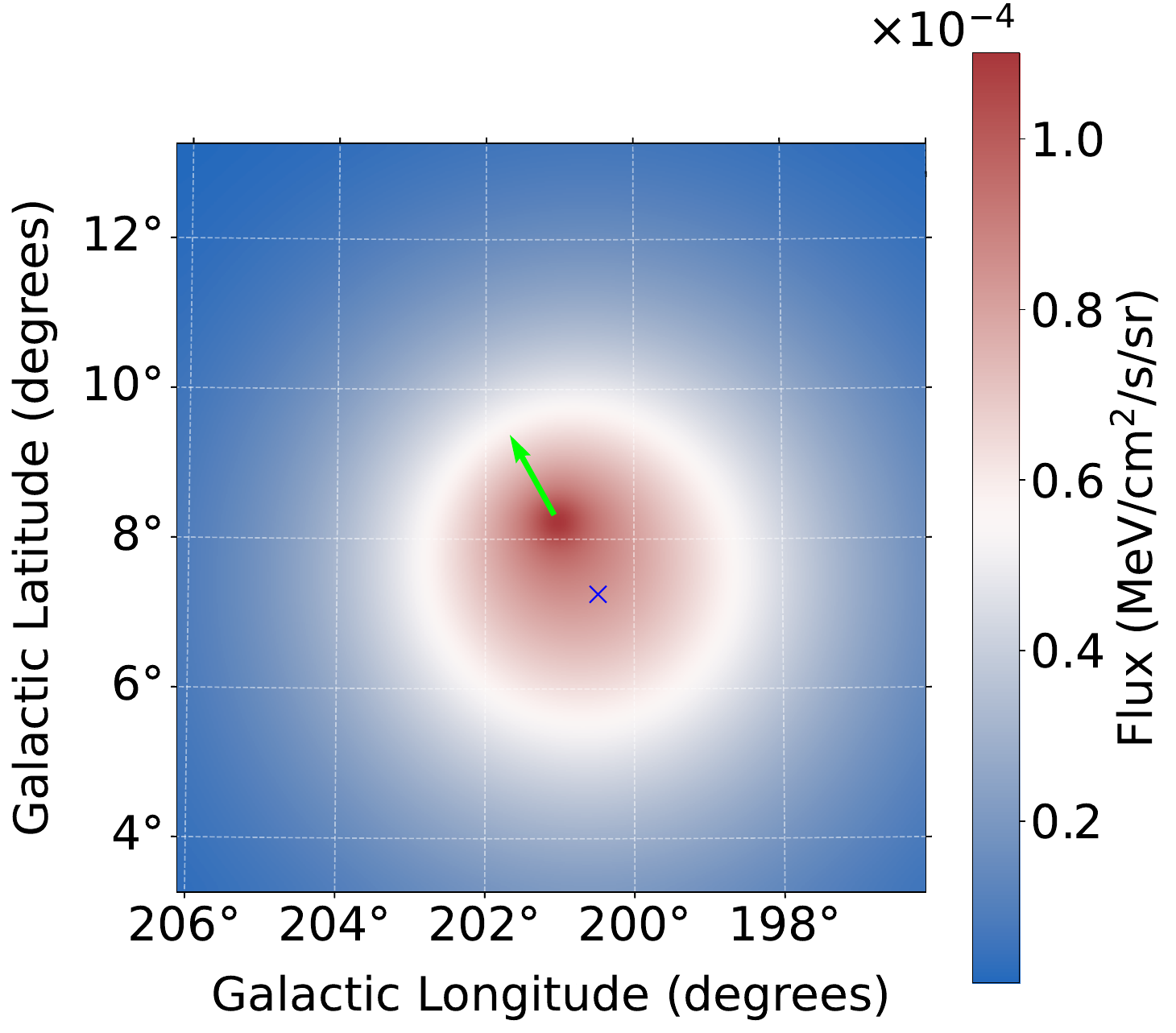}
  \caption{Synchrotron emission at 100 GHz}
  \label{fig:sync_pm_fixed_100GHz}
\end{subfigure}%
\begin{subfigure}{.49\textwidth}
  \centering  \includegraphics[width=.8\linewidth]{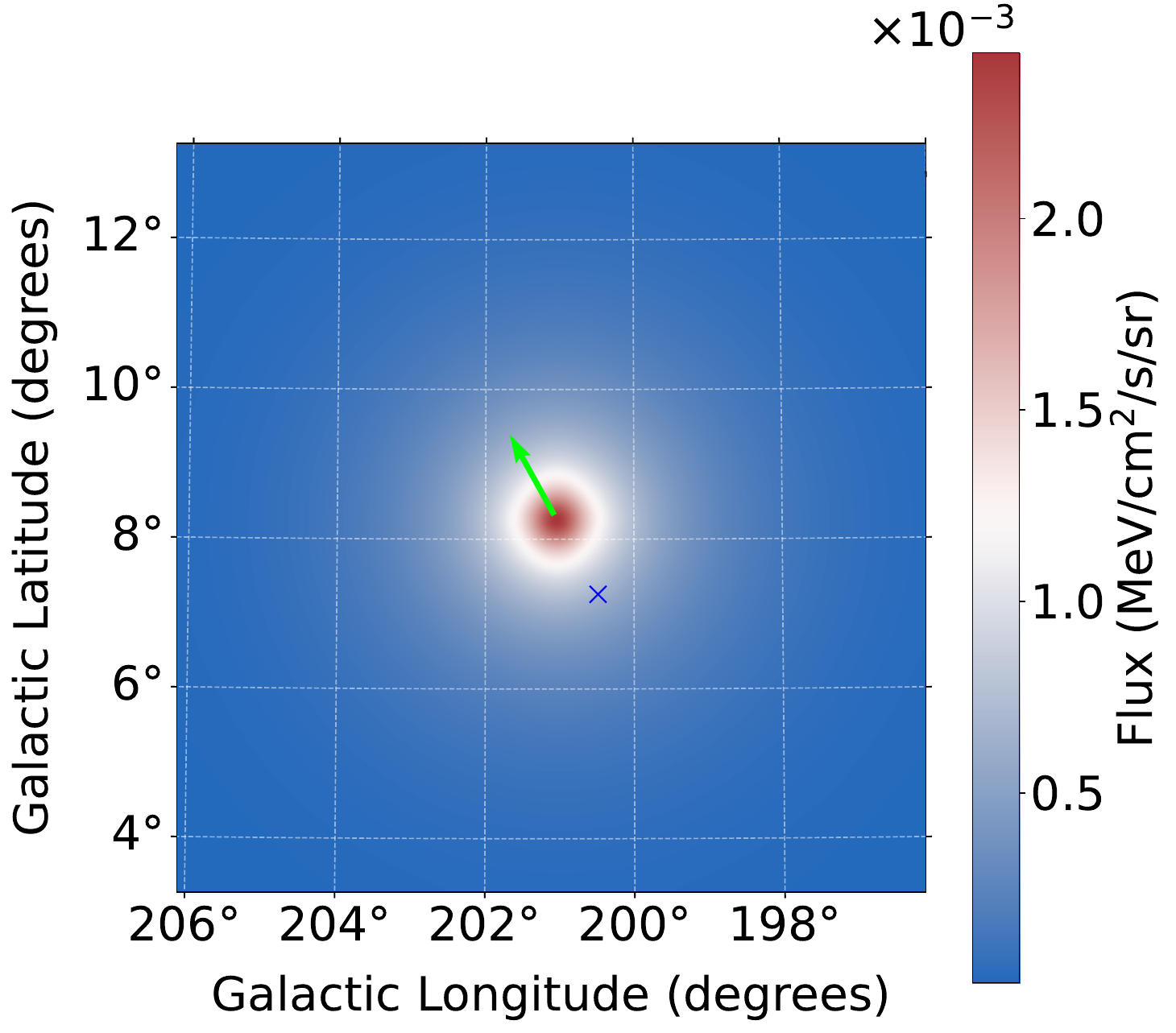}
  \caption{Synchrotron emission at 5 keV}
  \label{fig:sync_emission_fixed_5keV}
\end{subfigure}
    \caption{Intensity maps at different wavelengths for the region around Monogem pulsar taking into account the pulsar proper motion. The center of the SDZ stays at the birth location of the pulsar (Scenario C). The SDZ has a size of $(r_{1},r_{2})=(50,70)$~pc at current time. The pulsar true age is $t_{\rm age}= 0.9\tau_{\rm c}$. The magnetic field strength is $B=3~ \mu$G, and injection index $\gamma_{1}=2.2$. The lime vector shows the pulsar proper motion, and the blue cross marks the estimated pulsar birthplace. Asymmetry in the emissions due to the proper motion of the PWN and SDZ is significant for 10~GeV IC emission (a), and 100~GHz synchrotron emission (c). While higher symmetry is found in IC emission at 10~TeV (b), and synchrotron emission at 5~keV (d).}
\label{fig: pm_emission_fixedSDZ_map}
\end{figure*}

\hfill\break
Figure~\ref{fig: pm_emission_fixedSDZ_map} displays the intensity maps at 10 GeV and 10 TeV from IC emission and 100 GHz and 5 keV from synchrotron emission. Compared to Fig.~\ref{fig: pm_emission_map}, where the SDZ moves with the pulsar, we can see that the emission morphologies and intensities are comparable in every energy windows. Given the small proper motion of the pulsar perpendicular to the line of sight, the projected birthplace and the current pulsar location are only 1.2$^\circ$ apart in the sky. The impact of anisotropic diffusion due to the Monogem Ring over the pulsar's lifetime is minimal. In this scenario, the 10 TeV emission is approximately spherically symmetric around the pulsar, as predicted also in scenario B. This suggests that the asymmetric TeV halo observed by HAWC is unlikely to be caused by the spatial asymmetry of the diffusion coefficient introduced by the Monogem Ring. More realistic modeling of the diffusion bubble, taking into account the interaction between the PWN and the SNR, may help explain the observed lobed TeV emission. \\
\hfill\break
We present a comparison of synchrotron and IC spectra for scenarios A (stationary), B, and C in the bottom panel of Figure~\ref{fig: sync_IC}. The plot corresponds to an SDZ size of (50, 70)~pc, magnetic field $B = 3~\mu$G, injection spectral index $\gamma_{1} = 2.2$, and $t_{\rm age} = 0.9\tau_{\rm c}$. Consistent with the emission morphology discussed above, the emission spectra for scenarios B and C are indistinguishable due to the small proper motion of the pulsar. In \citet{Jóhannesson_2019}, the influence of the pulsar's proper motion and its interaction with Geminga's SDZ is explored, with a tangential proper motion estimated to be four times of that of Monogem. Their findings suggest that if Geminga's SDZ aligns with its birthplace, sub-keV to 10~keV synchrotron emission will be suppressed compared to a scenario where the SDZ tracks the pulsar’s motion. This configuration would also yield a distinctive IC emission tail at 10GeV.
\subsection{Positron Flux} \label{positron_result}
The PAMELA telescope observed a surplus in the CR positron-to-electron ratio at energies greater than 100 GeV, compared to the ratio predicted by the secondary production model \citep{Adriani_2008}. This observation was later confirmed by Fermi-LAT  \citep{Ackermann_2012}  and AMS-02 \citep{Aguilar_2013}, amongst others. Many authors have proposed various theories to explain this excess in the measured positron fraction, including nearby astrophysical leptonic sources \citep[e.g.,][]{Yuksel_2009,Dan_Hooper_2009}, and dark matter annihilation \citep[e.g.,][]{Ibe_2013,Dev_2014}. Geminga and Monogem are well-known nearby sources of $e^{\pm}$ that have emerged as potential sources of the positron excess. The TeV halos observed around these two pulsars provide strong evidence of leptonic particles escaping into the interstellar medium from these sources. However, the extension of the TeV halos also indicates that the $e^{\pm}$ are strongly confined, which suggests a reduced flux of positrons from these sources to Earth.

\begin{figure}
     \centering
     \begin{subfigure}[b]{0.5\textwidth}
        \centering
         \includegraphics[height=0.28\textheight]{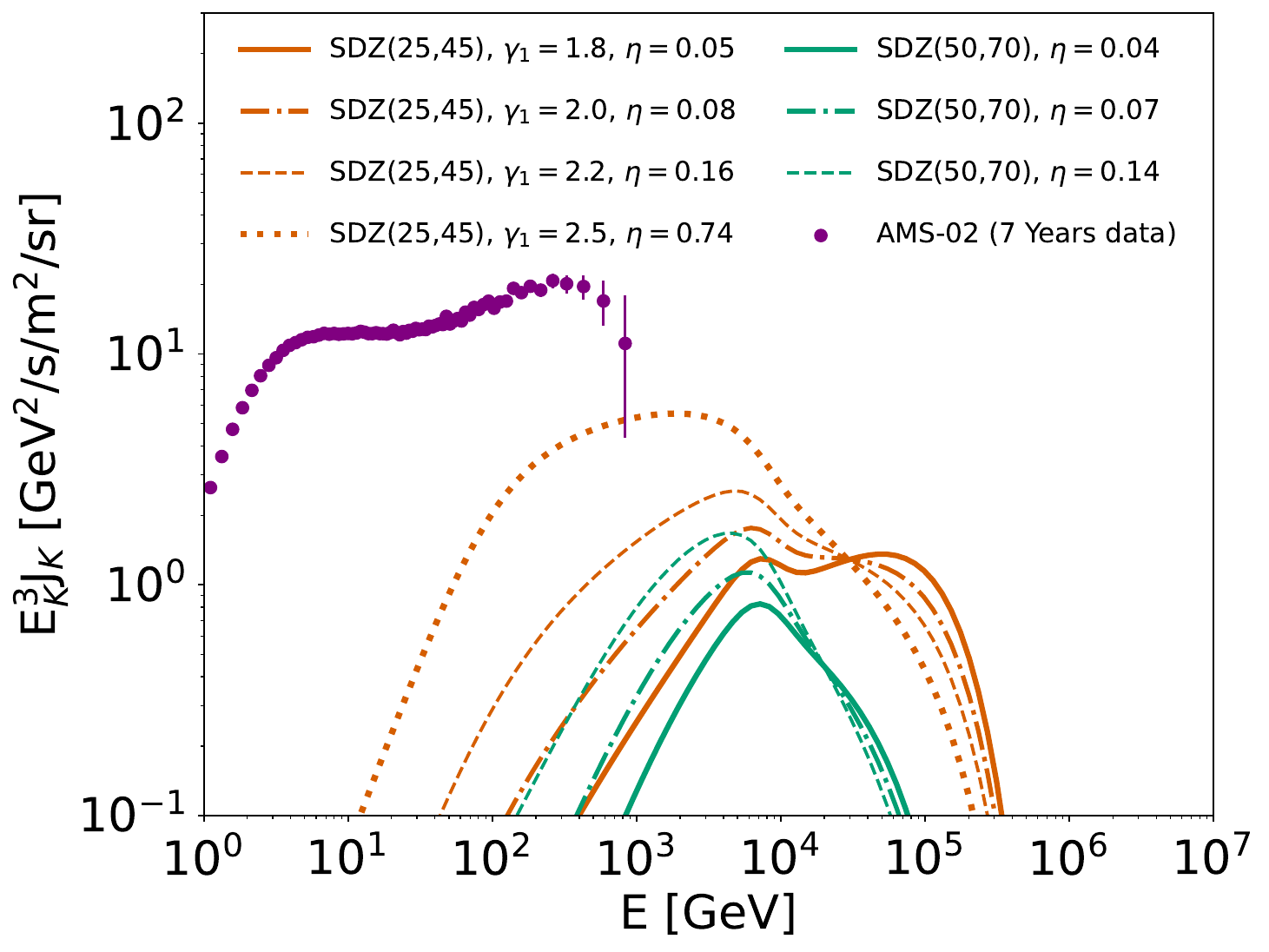}
%        \subcaption{}
     \end{subfigure}
     \hfill
     \begin{subfigure}[b]{0.5\textwidth}
         \centering
          \includegraphics[height=0.28\textheight]{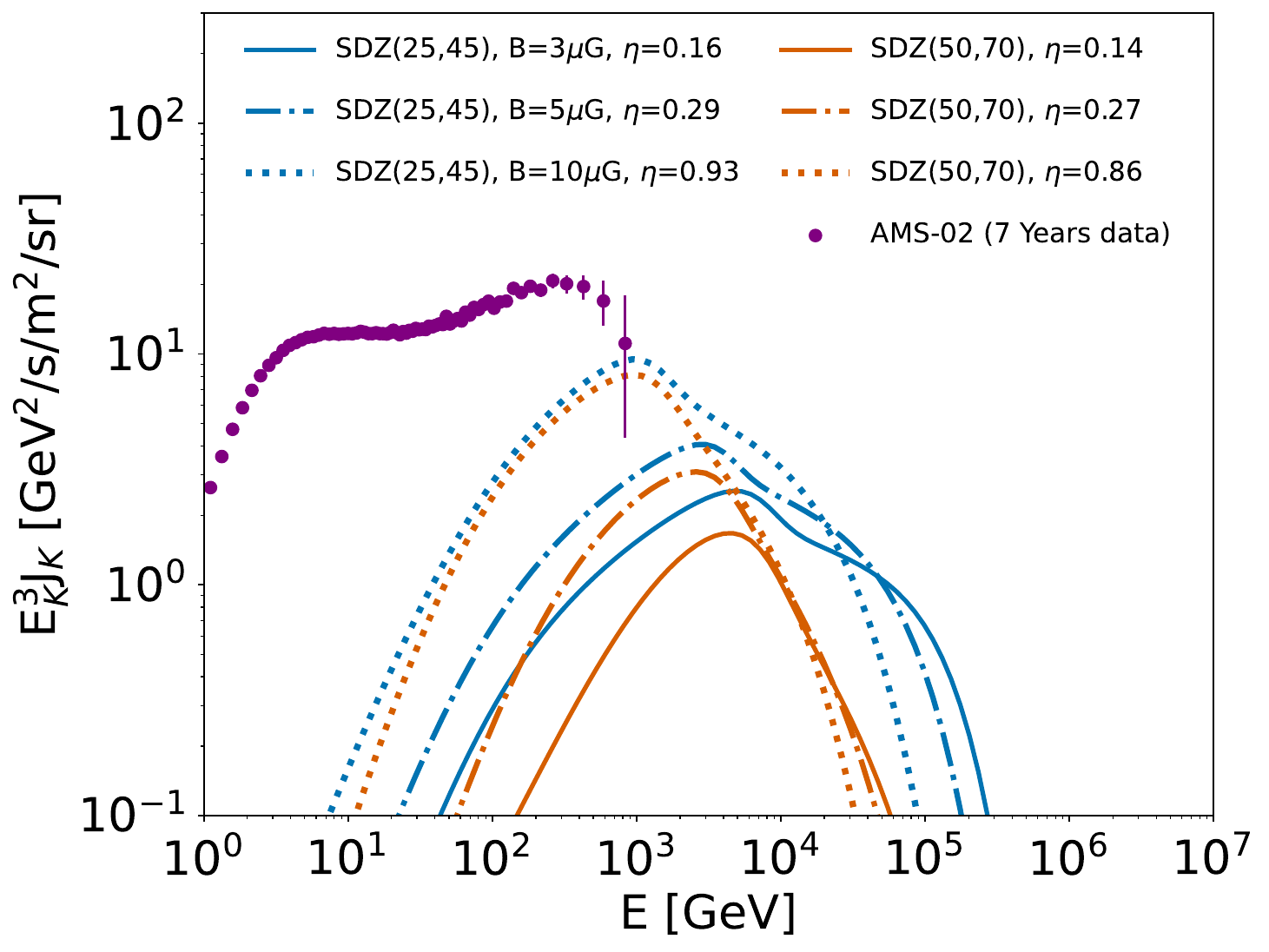}
%         \subcaption{}
     \end{subfigure}
     \caption{Positron flux from Monogem pulsar expected at Earth in comparison with AMS-02 7 years data. The pulsar true age is $t_{\rm age}=0.9 \tau_{\rm c}$ for all cases. \textbf{Top:} positron flux of varies injection index $\gamma_{1}=$ 1.8, 2.0, 2.2, 2.5, in scenario B. Magnetic field is 3~$\mu$G.  \textbf{Bottom:} positron flux of varies magnetic field B=3, 5, 10~$\mu$G, in scenario B. The injection index $\gamma_{1}=2.2$.}  
     \label{fig: positron_flux}
\end{figure}

\hfill\break
We have conducted a study on the primary positron spectra that are expected at Earth for different combinations of model parameters. In Fig. \ref{fig: positron_flux}, we show the data collected by AMS-2 over seven years and compare it to our prediction.

\hfill\break
Our study has found that the contribution of positrons with energy greater than 100 GeV from Monogem alone could vary from only a fraction of a percent to $\sim 20\%$ depending on the model parameter combinations. Increasing the size of SDZ results in a longer confinement time of positrons thus leading to lower positron flux contribution at the Earth. In the upper panel of Figure \ref{fig: positron_flux}, the softer the injection spectrum is, the higher the contribution to the positron flux is expected from Monogem. A $\sim 20\%$ contribution to $\sim 100$~GeV positron flux is expected for the case when injection spectra index $\gamma_{1}=$2.5, and SDZ size of (25, 45)~pc, with a required injection efficiency of 74\%.

\hfill\break
The high-energy positron spectra become softer due to synchrotron and IC losses. We have observed that the peak of the positron spectrum shifts from $\sim 5$~TeV to $\sim 1$~TeV as the magnetic field strength increases from 3~$\mu$G to 10~$\mu$G, because positrons lose their energy faster around the PWN. In the simulations, we have varied the magnetic field on the Galactic scale, and as a result, we expect the cutoff effect shown in the plot to be more dramatic than it would be in the actual case where only the magnetic field around the source is varied. Therefore, the positron flux estimated for a magnetic field larger than the Galactic average (3~$\mu$G) in the lower panel of Figure \ref{fig: positron_flux} should be taken as the lower limit.  Although the cases when $B=10~\mu$G are shown to contribute significantly to the positron excess, the combination of reasons disfavor these scenarios:
\begin{enumerate}
    \item the estimation is the lower limit and it contributes to almost all of the 1~TeV positron flux,
    \item observations of X-ray halo counterparts infer a magnetic field strength close to the Galactic average of 3~$\mu$G or lower, and
    \item the required injection efficiency of the Monogem pulsar to reproduce the observed TeV halo for a $10~\mu$G magnetic field is $\sim 90\%$, which is considerably high.
\end{enumerate}

\hfill\break
Despite this result, the pulsar hypothesis to explain the 100~GeV positron excess remains the most compelling one. Recent studies~\citep[e.g.,][]{Orusa:2021tts} have conducted sophisticated simulations of pulsar populations in the Galaxy, showing that nearby bright sources could account for the positron excess. These findings are consistent with previous analyses~\citep{Profumo:2018fmz, J_hannesson_2019, Schroer_2023, Martin_2022} of the Geminga pulsar, that show that the Geminga pulsar could contribute a significant proportion of the observed positron excess.
\section{Conclusions} \label{sec:style}
In this article, we utilized the GALPROP (v57) framework to calculate the expected spectrum and spatial distribution of interstellar diffuse gamma-ray, including IC and synchrotron emission, and positron spectra at Earth. We have demonstrated that HAWC's observations of the TeV halo surrounding PWN B0656+14 (Monogem) can be well explained by a two-zone diffusion model. Our modeling suggests that a diffusion bubble of an order of a few 10~pc in radius encompasses Monogem, where the diffusion coefficient experiences a suppression similar to that of the Geminga pulsar. These results imply that the slow diffusion of ultrarelativistic $e^{\pm}$ pairs around Monogem and Geminga may be due to the same mechanism.

\hfill\break
The proper motion of the Monogem pulsar introduces asymmetries in the GeV IC emission and the radio synchrotron emission around the pulsar's current location in both scenarios: whether the slow diffusion bubble is always centered at the pulsar or fixed at its birthplace. Our analysis shows that the non-thermal spectra and morphologies are indistinguishable between these two cases due to the small distance the pulsar has traveled perpendicular to the line of sight over its lifetime. Notably, both scenarios predict close to spherically symmetric 10 TeV emission, contrary to the asymmetrical emission morphology observed by HAWC. This suggests that the slow diffusion of pairs around Monogem may be due to additional factors beyond magnetic turbulence caused by the Monogem Ring or the propagation of the CRs themselves. Possible reasons could include pre-existing features of the diffusion coefficient in the surrounding ISM. Furthermore, realistic modeling that determines the exact location of the Monogem relative to the Monogem Ring, and their interactions, is necessary to understand the emission region. We note here that the TeV halo is expected to have synchrotron halo counterparts in the radio to X-ray energy range. However, because the projection of the Monogem PWN is embedded in the X-ray emission of the Monogem Ring, detecting the X-ray halo remains a challenge. 

\hfill\break
Based on comparisons between our models and the data, we obtained that the escape efficiency of ultrarelativistic $e^{\pm}$ from the PWN into the ISM is $\sim 5$--$30\%$ for an injection spectrum with $\gamma_{1}=2.2$, excluding the high magnetic field cases where $B=10~\mu$G due to several reasons mentioned in \ref{positron_result}. In this case, the corresponding GeV emission spectrum is around one order of magnitude lower than the upper limits obtained with Fermi-LAT \citep{Di_Mauro_2019} for a minimum SDZ size of (25, 45)~pc, and is close to the upper-limit when assuming a (50, 70)~pc size SDZ. 

\hfill\break
We showed that softer injection spectra (larger $\gamma_{1}$), lead to higher fluxes expected at GeV energies. An injection index of $\gamma_{1}>~2.5$ is challenging for an injection spectrum break energy at $E_{\rm b}=100$~GeV, because the predicted GeV emission for $\gamma_{1}=2.5$ remains just below the Fermi-LAT upper limit with a minimum SDZ size of (25, 45)~pc. We note that an improved analysis of Fermi-LAT observations could be used to constrain the properties of the SDZ further. Although it might be challenging to distinguish between the size of the diffusion bubble and the shape of the $e^{\pm}$ injection spectra, our predicted asymmetry in the spatial morphology of the GeV gamma-ray emission could help disentangle the Monogem pulsar from the diffuse gamma-ray background in Fermi-LAT observations.

\hfill\break
Moreover, we showed that the strength of the magnetic field has a significant impact on the $\sim$10--100~KeV synchrotron flux, and the required injection efficiency of the PWN to reproduce the TeV halo. Different diffusion coefficients are required to account for the varied cooling time of $e^{\pm}$ when assuming different magnetic field strength. With a diffusion coefficient suppression rate ($D_{\rm SDZ}/D_{\rm ISM}$) of 0.0027--0.0143, for a magnetic field of 3-10~$\mu$G, the HAWC surface profile was reconstructed. However, the high magnetic field cases are unlikely, when B=10~$\mu$G, because it results in a close to 100\% contribution to the observed $1$~TeV positron flux for our lower limit estimation, and requires a $\sim90\%$ injection efficiency of Monogem.

\hfill\break
It was claimed~\citep{HAWC_2017} that the $e^{\pm}$ produced by these pulsars may not contribute significantly to the locally measured positron flux if the diffusion coefficient in the local interstellar medium is similar to the value inferred from the Geminga and Monogem TeV halos. Here, in agreement with various previous studies \citep[e.g.,]{Profumo:2018fmz, Johannesson:2018bit}, we have shown in detail that if the diffusion is slow around the PWN for a few $10$~pc, then the expected contribution by Monogem could be at a fraction of a percent to $\sim$20\% level, which is not insignificant. We observed that a larger injection efficiency of the pulsar and a smaller size slow diffusion bubble would result in a higher contribution to the positron flux. Considering that there are many more objects like it within the inner $\sim 2$ kpc of the Solar system, pulsars may still be responsible for the positron excess. Indeed, recent studies~\cite[e.g.,][]{Orusa:2021tts} have demonstrated that a population of bright nearby pulsars could explain the AMS-02 positron data in detail. 

\hfill\break
Finally, we have shown that two-zone diffusion models can adequately explain multi-wavelength data from Monogem. A similar conclusion was obtained by~\citet{Porter_2017} in the case of Geminga. If slow-diffusion bubbles are a common feature of PWNe in our Galaxy, regions of the sky with high concentrations of these sources (e.g., the Galactic center) may have significantly higher cosmic-ray densities than those predicted by the standard diffusion model. These findings motivate further investigation of diffuse non-thermal radiation from the Galactic disk and bulge in which slow-diffusion bubbles wrap at least a fraction of the Galactic CR sources.

\section*{Acknowledgements}
We would like to express our gratitude to the GALRPOP team for making their code publicly available, and Troy Porter for his kind support in response to our inquiries. 

\hfill\break
OM was supported by the U.S. National Science Foundation under Grant No 2418730.
The work of SA was supported by JSPS/MEXT KAKENHI under grant numbers JP20H05850, JP20H05861, and JP24K07039.

\hfill\break
Our simulations were executed on the Dutch National supercomputer, {\it Snellius},\footnote{\url{https://visualization.surf.nl/snellius-virtual-tour/}} using a single node with 224 GB of memory and 128 CPUs. Each simulation took around $\sim 5$ hours with the computing node operating at total capacity.
\section*{Appendix}
\subsection*{A: GALPROP Setups}
We utilized GALPROP v57 to perform our calculations. Our simulations were conducted on a non-equidistant grid with high spatial resolution. The grid size varies based on a non-linear (tangent) transformation, as described in detail in \citet{Porter_2022}. This grid setup allows for maximum resolution at a specific point in 3D space, decreasing at greater distances and optimizing simulation memory usage and speed. The grid size at the current location of the Monogem pulsar is 2~pc, while at a reference distance of 700~pc, it increases to 100~pc. The resolution is roughly uniform in the $\sim 60$~pc region from the source, encompassing the entire evolution of the Monogem PWN. To solve the transport equation in the time domain, we used a time step of 50 years, which is sufficient to capture the pairs' energy loss. We ensured that the number of time steps was adjusted to cover the entire system's lifetime based on the different true pulsar ages we tested.

\hfill\break
The GALPROP coordinate system is right-handed, centering at the Galactic center, with the Sun locating at $(X_{s},\,Y_{s},\,Z_{s}) = (8.50,\,0,\,0)$~kpc.
The Monogem pulsar is positioned at the Galactic coordinates  $(l,\,b)=(201.1^{\circ},\,8.3^{\circ})$, which corresponds to $(X,\,Y,\,Z) = (8.766,\,0.103,\, 0.042)$~kpc in GALPROP coordinates. Its proper motion is measured to be $\mu_{\alpha}\cos(\delta)=44.16$ mas/yr, and $\mu_{\delta}=-2.43$ mas/yr \citep{Hobbs_2005}. Assuming that the pulsar velocity is constant, the line of sight velocity is zero, and the pulsar age is 99~kyr (corresponds to $\tau_{0}=0.1\tau_{\rm c}$), the Monogem pulsar moved roughly from $(X_{0},\,Y_{0},\,Z_{0}) = (8.768,\,0.100,\,0.036)$ kpc with a velocity of $(V,\,U, \,W)= (-17.53,\,24.34,\,52.09)$~km~s$^{-1}$ over its lifetime, where $V$, $U$, and $W$ are velocities along the $X$, $Y$ and $Z$ axis respectively.

\hfill\break
We used the ``Sun\_ASS\_RING\_2''  Galactic magnetic field model and the R12 interstellar radiation field model to calculate the synchrotron and inverse Compton emissions. The magnetic field model is based on \citet{Sun_2008} which consists of regular and random fields. At the location of the Sun, the regular field is $2\;\mu$G, and the random field is $3\;\mu$G. The magnetic field strength also has an axisymmetric spiral component with reversals in rings. However, it is not flexible enough to consider the local variance of the magnetic field around the PWN due to magnetic turbulence. To test how the magnetic field strength affects the non-thermal emission, we compare cases where the random field is $3\;\mu$G, $5\;\mu$G, and $10\;\mu$G at the location of the Sun. It is important to note that even though this modification impacts the magnetic field on a Galactic scale, it is still valuable for estimating the impact on the non-thermal emission around the source. In section \ref{positron_result}, we will show that the additional synchrotron loss during propagation, caused by an enhanced magnetic field, leads to some suppression in our positron flux estimation at Earth.
\section*{Data Availability}
The data underlying this article will be shared on reasonable request to the corresponding author.


\bibliographystyle{mnras}
\bibliography{mmmonogem} % if your bibtex file is called example.bib

%\begin{thebibliography}{99}
%\end{thebibliography}

%%%%%%%%%%%%%%%%%%%%%%%%%%%%%%%%%%%%%%%%%%%%%%%%%%

%%%%%%%%%%%%%%%%% APPENDICES %%%%%%%%%%%%%%%%%%%%%

\appendix



%%%%%%%%%%%%%%%%%%%%%%%%%%%%%%%%%%%%%%%%%%%%%%%%%%

% Don't change these lines
\bsp	% typesetting comment
\label{lastpage}
\end{document}